\documentclass{IEEEtran}
\usepackage{graphicx}
\UseRawInputEncoding
\usepackage[cmex10]{amsmath}
\usepackage{hyperref}
\begin{document}

%\title{A Sample Article Using IEEEtran.cls\\ for IEEE Journals and Transactions}
\title{New Closed-Form ASER Expressions for Dual-Hop Mixed THz-RF Cooperative Relay Networks}

%\author{X~~~~~~~~~~~~~Y~~~~~~~~~~~~~~
       %~Z}% <-this % stops a space}
\author{Soumendu Das, Nagendra Kumar, Dharmendra Dixit
        % <-this % stops a space
\thanks{Soumendu Das and Nagendra Kumar are with the Department of Electronics and Communication Engineering, National Institute of Technology Jamshedpur, Jharkhand-831014, India (e-mail: 2021rsec009@nitjsr.ac.in, kumar.nagendra86@gmail.com

Dharmendra Dixit is with the Department of Electronics Engineering, Rajkiya Engineering College, Sonbhadra, U.P. India (ddixit@recsonbhadra.ac.in)}
}

\maketitle

%====================================================================================================================
% # I. Abstract #
%====================================================================================================================

\begin{abstract}
 In this paper, we consider a dual-hop mixed THz-RF system model for backhaul-fronthaul applications where the link between source and destination is established only through the relay node in which decode-and-forward relaying protocol is used. The THz link suffers from the joint impact of antenna misalignment and stochastic characteristics of wireless channels, including the effect of environmental conditions such as pressure, humidity, and temperature. The envelope of THz link in the first hop follows a generalized $\alpha-\mu$ distribution, and for the RF end, the Nakagami-$m$ distribution is considered. In this context, we obtain new closed-form expressions of the cumulative density function and the moment-generating function of the end-to-end signal-to-noise ratio. Further, we derive the average symbol error rate expressions for coherent rectangular quadrature amplitude modulation (RQAM) and coherent hexagonal QAM (HQAM), as well as the non-coherent modulation scheme. The asymptotic behavior is also discussed to examine the system's diversity. Furthermore, the impact of several parameters, such as fading coefficients of individual links and antenna misalignment, as well as the distance between nodes, are also highlighted in the system's performance. Moreover, Monte Carlo simulations are used to validate the presented analytical framework. Finally, the presented numerical insights aid in the extraction of practical design principles.

\end{abstract}

% \IEEEoverridecommandlockouts

\IEEEpeerreviewmaketitle

%====================================================================================================================
% # II. Introduction #
%====================================================================================================================

\section{Introduction} \label{introduction}
% Introduction (1/7) : Regarding What is THz, Why THz, previous work
\IEEEPARstart{O}{ver} the last few years, the number of wireless devices are increased at an expeditious rate and thus the significant accretion of wireless data traffic has encouraged the investigation of suitable regimes in the higher frequency bands of unused radio spectrum such as the terahertz (THz) band\cite{8901159} to meet the escalating requirements of higher data rates with negligible latency. Unlicensed THz spectrum is cost-effective and theoretically competent to provide extremely high bandwidth \cite{8387218,AKYILDIZ201416,8387219} due to its existence in the higher region of the spectrum (0.1-10 THz) \cite{7765354}. Thus, THz band is appropriate for wireless fiber extenders in backhaul link particularly when fiber optic link is burdensome to establish due to geographical challenges. 
% Introduction (2/7) : Challenges in THz : Molecular absorption
Despite these advantages, the THz spectrum immensely suffers from severe attenuation due to molecular absorption \cite{7061628}. When a signal with such a high frequency interacts with molecules of air medium, the energy of the signal is absorbed mostly due to the water vapor molecules present in the atmosphere \cite{8387210}. Generally, $\alpha-\mu$ distribution is used to model THz communication link because of its versatile statistical behavior which includes several well-known channel models such as Rayleigh, Nakagami-$m$, Weibull, Hoyt, Rice, Gamma, Chi-square, etc. \cite{4067122}.

\subsection{Related Literature} \label{LiteratureSurvey}
In literature, a novel THz band propagation model is discussed in \cite{5995306}. In order to evaluate the performance of THz link, a path loss model due to molecular absorption in the $275-400$ GHz band is considered in \cite{8568124} and the channel capacity of THz link is calculated in \cite{8417891}. Whereas, absorption loss due to components of propagation medium such as air, water, and natural gas is presented in \cite{akkacs2016terahertz}.
Likewise in \cite{6998944}, authors have discussed a multi-ray THz model.

% Introduction (3/7) : challenges in THz : Hardware misalignment
Performance of THz system can degrade severely due to hardware imperfections like in-phase and quadrature imbalance, minute misalignment of highly directional transceiver antennae\cite{8932597}, etc. Therefore, an extensive investigation is mandatory in modeling the particularities of THz channel as it is relatively new to explore. The impact of antenna misalignment is investigated in \cite{6205849} and \cite{7579223}. The influence of highly directive antennae in the THz band is studied in \cite{6205849}. In \cite{8610080}, authors have investigated the joint impact of antenna misalignment and fading in THz-wireless systems in terms of outage probability and channel capacity.

% Introduction (4/7) : why not THz in second hop. paper [4] rayleigh. Nakagami-m is more generalised
A multipath model for THz channel is discussed in \cite{8088634} and \cite{5723718} considering Nakagami-$m$ fading distribution for line-of-sight as well as non-line-of-sight conditions and experimental results show the existence of shadowing in $300$ GHz band \cite{5723718}. As a consequence, THz spectrum is questionable for fronthaul due to its high sensitivity to blockage. Thus, the second hop is assumed to work in the RF spectrum.

% Introduction (5/7) : relaying Scheme
Further to improve the quality of communication, various relaying schemes can be employed at the relay node such as decode-and-forward (DF), amplify-and-forward (AF), compress-and-forward (CF), etc. Although, AF is mostly preferred as it is cheap and the complexity of signal processing is very low. In spite of the complexity of decoding, DF can be used as relaying protocol due to its better performance than AF. In \cite{9937036} and \cite{9449062}, authors have modeled a dual-hop THz system considering AF and DF relaying protocol, respectively, and checked its suitability for backhaul communications. In \cite{9885233}, authors have modeled a multihop system operating over THz frequency range while in \cite{6630485}, the performance of RF band relaying system is observed. In \cite{9492775}, authors have modeled a dual-hop mixed THz-RF model and calculated average symbol error rate (ASER) expressions for differential binary phase shift keying (DBPSK), $M$-ary pulse amplitude modulation (PAM) and non-return-to-zero (NRZ) on-off keying (OOK) modulation schemes.

% Introduction (6/7) : modulation scheme, sir's papers
 %and cross QAM (XQAM) \cite{1545878}. 
Further, limited bandwidth is optimally utilized to achieve enhancement in data rates by an adaptive selection of higher-order quadrature amplitude modulation (QAM) schemes such as rectangular QAM (RQAM) \cite{1632093}, squared QAM (SQAM) and hexagonal QAM (HQAM)\cite{7480856}. Amongst these RQAM is the most flexible modulation scheme as it includes special cases such as SQAM, quadrature phase shift keying (QPSK), multi-level amplitude shift keying (M-ASK), and orthogonal binary frequency shift keying (OBFSK). It has a lot of applications such as asymmetric subscriber loops, high-speed mobile communication, etc.
% In \cite{7544579}, authors have performed the exact average symbol error rate analysis for rectangular QAM modulation in a two-way relay network. Hexagonal QAM consists of optimal constellation points such that densest two dimensional (2D) constellation of hexagonal lattice shape can be achieved and thus it is capable to provide significant signal-to-noise ratio (SNR) gain over other modulation schemes \cite{8053787},\cite{8125102},\cite{SINGYA2018337}. 
Several authors have discussed the performance of higher-order QAM schemes for several system and channel models \cite{9857864,DIXIT2021153883, 9350626, 9382012,9785975,singh2022aser}. To the best of our knowledge, the ASER performance of practically useful QAM schemes for dual-hop mixed THz-RF cooperative relay networks is not explored yet.

%However, for odd number of constellation points, RQAM is not a good choice and an optimum XQAM constellation is preferred due to its lower peak and average power. The XQAM constellation is formed by modifying the RQAM constellation with the removal of outer corner points and arranging them in such a manner that the peak and average power of the constellation is reduced \cite{1545878}. XQAM is used in various practical systems such as very high bit-rate digital subscriber line (VDSL), asymmetric digital subscriber line (ADSL), digital video broadcasting-cable (DVB-C) \cite{zhang2010exact}.

\subsection{Motivations} \label{motivations}
With the move towards 6G and beyond wireless technologies, ultra-high data transmission rates, extremely large bandwidth, and very low latency are essential for multimedia transmission over wireless channels. 3rd generation partnership project (3GPP) Release-14 (LTE-advanced) and Release-15 (5G-New radio) have suggested the usage of higher order QAM, such as 256-QAM and 1024-QAM, for different high data rate applications \cite{3gppR15}. Taking the above discussions into consideration, the mixed THz-RF based relay networks with QAM schemes become a potential technique to meet these demands. Although, in literature, dual-hop mixed THz-RF based cooperative relay networks have not been studied yet to analyze the ASER performance of practically important coherent QAM schemes. It motivates us to study and analyze the considered system, which will be of immense interest in 6G and beyond for the system designers.
%%%%%==== Contributions
\subsection{Contributions} \label{contributions} % Steps written in own language
For the THz wireless fiber extender system, we study a suitable and generalized system and channel model that takes into account many design factors as well as their interactions. In this paper, we consider a dual-hop mixed THz-RF based DF relay network considering $\alpha-\mu$ and Nakagami-$m$ distribution for the THz and RF links, respectively. The main contributions of this manuscript are as follows:
\begin{enumerate}
    \item We obtain a new closed-form expression for the cumulative density function (CDF) of the end-to-end signal-to-noise ratio (SNR).
    \item A new moment-generating function (MGF) of end-to-end SNR is derived using the obtained CDF expression.
    % \item The closed-form expressions of the outage probability, and moment-generating function (MGF) are also derived.
    \item The generalized ASER expressions of coherent RQAM and HQAM schemes are obtained using the well-known CDF-based approach.
    \item The generalized ASER expression of non-coherent FSK (NCFSK) is also obtained using the MGF-based approach. 
    \item In addition, the asymptotic expression of ASER for RQAM is presented.
    \item The diversity order is obtained with the aid of the derived asymptotic ASER expression.
\end{enumerate}

% Notation Section
\subsection{Notations} \label{notations}
Some notations that are used in the paper are as follows. $\Gamma(\cdot)$ and $\Gamma(\cdot, \cdot)$ represent the complete gamma function \cite[eq. (8.310.1)]{gradshteyn2014table} and upper incomplete gamma function \cite[eq. (8.350.2)]{gradshteyn2014table}, respectively,  $_{1}F_{1}(\cdot; \cdot; \cdot)$ represents the confluent hypergeometric function of first kind \cite[eq. (9.210.1)]{gradshteyn2014table}, ${}_pF_{q}\left(a_1,\ldots,a_p;b_1,\ldots,b_q;z\right)$ represents generalized hypergeometric function \cite[eq. (7.2.3)]{prudnikov}, $\binom{\cdot}{\cdot}$ represents the binomial coefficient, $G_{p, q}^{m, n}\left[z \left|\begin{array}{c}
    a_{1}, \ldots, a_{p} \\
    b_{1}, \ldots, b_{q}
    \end{array}\right.\right]$ represents the Meijer-G function \cite[eq. (9.301)]{gradshteyn2014table}, and Fox-H function \cite[eq. (17)]{SRIVASTAVA1979191} is represented by
$H_{p, q}^{m, n}\left[z \left|\begin{array}{c}
    (a_{1}, A_{1}), \ldots,(a_{p}, A_{p}) \\
    (b_{1}, B_{1}), \ldots,(b_{q}, B_{q})
    \end{array}\right.\right]$. Further, multivariate Fox-H function \cite[eq. (A.1)]{Mathai2010} can be expressed as
    \scalebox{0.8}[1]{
    $H_{p,q:p_1,q_1;\ldots;p_r,q_r}^{0,n:m_1,n_1;\ldots;m_r,n_r}\left[\left.\begin{array}{cc}
        z_1 \\
        \vdots \\
        z_r
    \end{array}\right|\begin{array}{cc}
         \left(a;\alpha_1,\ldots,\alpha_r\right)_{1:p}:\left(c,\gamma\right)_{1:p_1};\ldots;\left(c,\gamma\right)_{1:p_r}  \\
         \left(b;\beta_1,\ldots,\beta_r\right)_{1:q}:\left(d,\delta\right)_{1:q_1};\ldots;\left(d,\delta\right)_{1:q_r} 
    \end{array}\right].$}
\subsection{Organization of the paper} \label{organizationOfThePaper} % to analyze the diversity order
The rest of the paper is organized as follows: In Section \ref{systemAndChannelModel}, we discuss the system and channel model of relay assisted THz-RF system. In Section \ref{performanceAnalysis}, we analyze the performance of the proposed system model. In subsections~\ref{CDFExpressionForE2ESNR}, the expression for the CDF of end-to-end SNR is presented. In \ref{MGFExpressionForE2ESNR}, a new closed-form expression for the MGF of end-to-end SNR is derived. In subsection~\ref{ExacterrorRateExpressions}, ASER expressions for coherent QAM schemes and non-coherent FSK modulation schemes are derived. Moreover, asymptotic analysis of RQAM modulation scheme is also presented in subsection~\ref{AsymptoticerrorRateExpressions}. In Section \ref{numericalAndSimulationResults}, simulation and theoretical results are compared. Finally, the paper is concluded in Section \ref{conclusion}.

%====================================================================================================================
% # III. System and Channel Model/System model #
%====================================================================================================================

\section{System and channel model} \label{systemAndChannelModel}

% =========== System Model ================
\subsection{System model}  \label{systemModel}
\begin{figure}[t!] %!t
\centering
\includegraphics[width=0.5\textwidth]{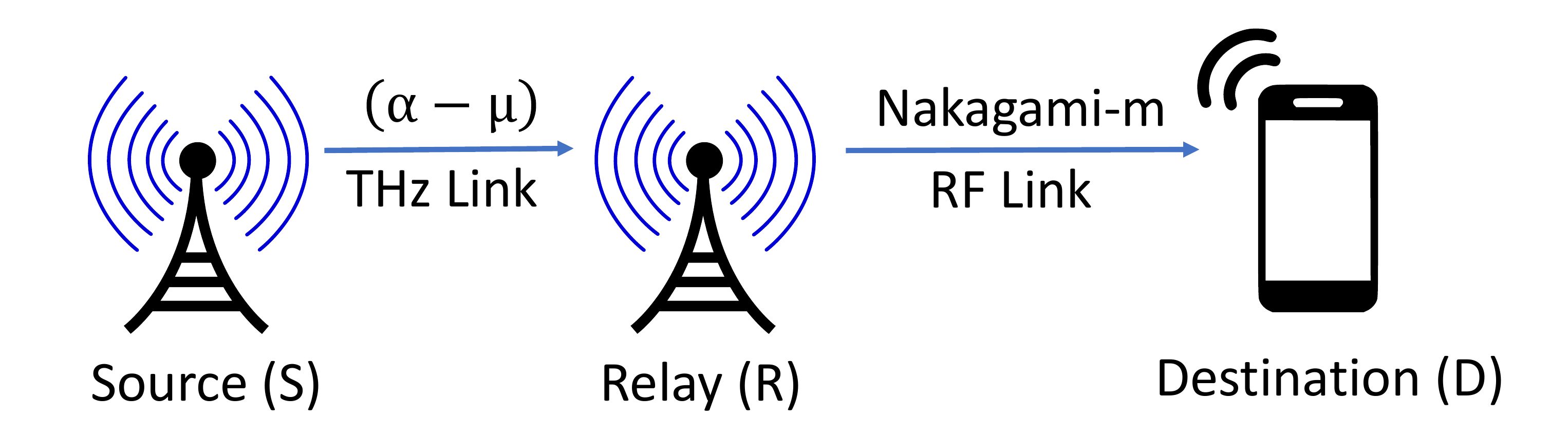}
\caption{THz-RF System Model}
\label{System Model Figure}
\end{figure}

We consider a dual-hop mixed THz-RF wireless system model consisting of a source node (S), a destination node (D), and a relay node (R), as shown in Fig. \ref{System Model Figure}. It is assumed that the direct link between S and D is extremely weak, so it can be neglected. Therefore, the communication link between S and, D is established through R only, where it uses DF relaying technique. Therefore, the overall downlink communication is accomplished in two-time slots. In the first time slot, the symbol is transmitted from S through the S-R channel and received by R. In the next time slot, R decodes the received symbol. If the symbol is found to be correct, the relay re-encodes the symbol to transmit it over R-D link. Finally, D decodes the symbol from the received signal.

% =========== Channel Model ================
\subsection{Channel model}  \label{channelModel}

% =========== THz Link ================
\subsubsection{THz link (hop 1)}  \label{THzLinkHop1}
In the first hop, symbols are transmitted from S over the S-R link. Thus, the signal received by R can be mathematically expressed as
\begin{align} \label{hop1_eqn}
   y_r=\sqrt{P_s}\,h_{sr}\,x_s + w_r,
\end{align}
where $P_s$ is the power of the signal transmitted from S, $h_{sr}$ is the channel coefficient of S-R link, $x_s$ is the transmitted symbol having unit energy, and $w_r$ is the additive white Gaussian noise (AWGN) with zero mean and variance $N_1$. Further, the channel coefficient $h_{sr}$ can be represented as
\begin{align} \label{channel_coeff_1_eqn}
   h_{sr} = h_{d1}  h_{a1}  h_{f1}  h_{p1},
\end{align}
% ----------------- Free pace Path Loss equation---------------
where $h_{d1}$ represents the deterministic free space path-loss which can be defined using Friis equation\cite{balanis2011modern} as

% free space path loss equation
\begin{align} \label{fspl1_eqn}
    h_{d1}=\frac{c \sqrt{G_{ts}\, G_{rr}}}{4 \pi f_{sr}} \, d_{sr}^{-\left(0.5\,\eta_1\right)},
\end{align}
wherein $c$ is the speed of light in free space, $f_{sr}$ is the frequency of the signal transmitted by S, $d_{sr}$ is the distance between S and R, $G_{ts}$ is the transmitter antenna gain of node S and $G_{rr}$ is the receiver antenna gain of node R. For free space or air medium without shadowing, path loss exponent $\eta_1$ is considered to be 2. The value of $\eta_1$ increases in urban regions, and if shadowing is present due to a large factory or high-rise buildings, $\eta_1$ increases further.

% ----------------- Molecular absorption loss equation---------------

The deterministic loss due to atmospheric molecular absorption, $h_{a1}$, in THz frequency range is modeled by using Buck's equation \cite{alduchov1996improved} as
\begin{align} \label{molabs1_eqn}
    h_{a1}=\exp\left[-0.5\, \kappa_{\alpha}\left(f_{sr}, T, P, \psi\right)  d_{sr} \right],
\end{align}
where $\kappa_{\alpha}$ is the absorption coefficient \cite[eq. (8)]{9492775} and it depends on atmospheric parameters like absolute temperature (T), pressure (P), and relative humidity ($\psi$) present in the air. 

% ---------------  PDF and CDF of RV ( short term fading + multipath = alpha-mu distribution )---------
The envelope of multipath fading $h_{f1}$ in THz link is assumed to follow generalized $\alpha-\mu$ distribution described by fading parameter, $\alpha$ ($\alpha>0$), $\alpha$-root mean value, $\Omega$, and the normalized variance, $\mu$ ($\mu \geq \frac{1}{2}$), of the fading channel envelope. The probability density function (PDF) of $h_{f1}$ is represented with the aid of \cite[eq. (1)]{4067122} as

% Reference: https://ieeexplore.ieee.org/document/7429677
\begin{align} \label{alpha_mu_pdf}
    f_{h_{f1}}(x)=\frac{\alpha \mu^{\mu} x^{\alpha \mu-1}}{\Omega^{\alpha \mu}\, \Gamma(\mu)} \exp \left(-\frac{\mu \, x^{\alpha}}{\Omega^{\alpha}}\right).
\end{align}
%Whereas, the CDF expression of $h_{f1}$ is obtained by transforming the upper incomplete gamma function present in \cite[eq. (8)]{4067122} into its equivalent summation form by making use of \cite[eq. (8.352.2)]{gradshteyn2014table} as
% \textcolor{red}{With the  \cite[eq. (8.310.1)]{gradshteyn2014table}
Whereas, the CDF expression of $h_{f1}$ is obtained by integrating (\ref{alpha_mu_pdf}) with the aid of \cite[eq. (4.1)]{edition2002probability} and further, the integration can be expressed in terms of upper incomplete gamma function \cite[eq. (8.352.2)]{gradshteyn2014table} as

\begin{align} \label{alpha_mu_cdf}
    F_{h_{f1}}(x) = 1 - \frac{1}{\Gamma(\mu)} \Gamma\left( \mu, \frac{\mu x^{\alpha}}{\Omega^\alpha}\right).
\end{align}
Further, $h_{p1}$ represents the fading coefficient of pointing error due to antenna misalignment between S and R. Thus, it can be modeled as \cite{4267802}.
% ---------------- PDF of RV (Fading due to antenna misalignment and pointing error) ------------------------

\begin{align} \label{channel_coeff_1_random_1_eqn}
    f_{h_{p1}}(x)=\frac{\phi \, x^{\phi-1}}{S_0^{\phi}} , \quad 0 \leq x \leq S_{0}
\end{align}
where $S_0$ is the fraction of power collected at the receiver when transmitter and receiver antennas are fully aligned, and $\phi$ is the ratio of the square of equivalent beam radius $w_{z_{eq}}$ and standard deviation of pointing error displacement at the receiver $\sigma_s^2$. The value of $S_0$ and $\phi$ can be calculated as, $S_0=\left[erf\left(\nu\right)\right]^2$ and $\phi=\frac{w_{z_{eq}}^2}{4\sigma_s^2}$, where, $\nu$ and $w_{z_{eq}}$ are calculated from the radius of the detector aperture and beam footprint radius $w_z$ at a transmission distance $z$ as mentioned in \cite[(9)]{4267802}. We consider $h_{1}=h_{f1}\, h_{p1}$, thus, the 
PDF of $h_{1}$ can be evaluated with the aid of \cite{edition2002probability} as

% Hop 1 overall PDF equation
\begin{equation} \label{hop1_pdf}
    f_{h_1}(x)=\left[\frac{\phi \, \mu^{\left(\frac{\phi}{\alpha}\right)} x^{\left(\phi-1\right)}}{ (\Omega S_{0})^{\phi} \, \Gamma(\mu)} \right] \, \Gamma\left(\frac{\alpha \mu-\phi}{\alpha}, \frac{\mu x^{\alpha}}{(\Omega S_{0})^{\alpha}} \right).
\end{equation}
The CDF of $h_1$ can be calculated using (\ref{alpha_mu_cdf}) and (\ref{channel_coeff_1_random_1_eqn}) with the aid of \cite{edition2002probability} as expressed in (\ref{hop1_cdf}). The calculation of (\ref{hop1_cdf}) is elaborated in subsection~\ref{CDFISol} of the appendix.
% Hop 1 overall CDF equation
\begin{equation} \label{hop1_cdf}
    F_{h_{1}}(x) = 1 - \frac{1}{\Gamma\left(\mu\right)}\frac{\phi}{\alpha}G_{2,3}^{3,0}\left[ \frac{\mu x^\alpha}{\Omega^\alpha S_0^\alpha} \left| \begin{array}{c}
    1,\frac{\phi}{\alpha}+1 \\
    \mu,0,\frac{\phi}{\alpha}
    \end{array} \right.\right],
\end{equation}
\newline

% ======================================================= 
%                         RF Link                       #
% =======================================================
\subsubsection{RF link (hop 2)}  \label{RFLinkHop2}
In the DF relaying technique, node R decodes the received symbol of the first time slot. If the received symbol is found to be error-free, node R re-encodes the decoded symbol and transmits it through R-D link in the second time slot. The received signal at the destination can be expressed as
\begin{align} \label{hop2_eqn}
   y_{d}=\sqrt{P_r}\,h_{rd}\,\tilde{x}_r + w_{d},
\end{align}
where $\tilde{x}_r$ and $P_{r}$ are the re-encoded unit energy symbol transmitted from R and transmitted power at R, respectively. $h_{rd}$ is the channel coefficient of the second hop. $w_d$ is the AWGN with zero mean and variance $N_2$. In case of incorrect reception at R, no symbol is transmitted further. Also, similar to THz link, $h_{rd}$ can be represented in the form of deterministic free space path gain, $h_{d2}$, and fading channel coefficient, $h_{f2}$, as 
\begin{align} \label{channel_coeff_2_eqn}
   h_{rd} = h_{d2} h_{f2},
\end{align}
where the deterministic coefficient for free space path-loss, $h_{d2}$, can be modeled similar to (\ref{fspl1_eqn}) as

\begin{align} \label{hop2_deterministic}
    h_{d2}=\frac{c \, \sqrt{G_{tr}\, G_{rd}}}{4 \pi f_{rd}} d_{rd}^{-\left(0.5\,\eta_1\right)},
\end{align}
where $G_{tr}$ is the transmitter antenna gain of node R and $G_{rd}$ is the receiver antenna gain of node D. $f_{rd}$ and $d_{rd}$ are the frequency of the signal used in RF link and the distance it between nodes R and D, respectively. The envelope of $h_{f2}$ is assumed to follow the Nakagami-$m$ distribution. Thus, the PDF, and the CDF expressions of $h_{f2}$, can be given as, respectively \cite{NAKAGAMI19603}

\begin{align} \label{hop2_pdf}
    f_{h_{f2}}(x)=\frac{2 m^{m} x^{2 m-1}}{(\Omega_{m})^{m} \Gamma(m)} \exp\left(-\frac{m}{\Omega_m}x^{2}\right),
\end{align}
\begin{align} \label{hop2_cdf}
    F_{h_{f2}}(x)=1-\frac{1}{\Gamma(m)} \Gamma\left(m, \frac{m }{\Omega_m}x^{2}\right),
\end{align}
where $m$ is shape parameter and $\Omega_m$ represents spread controlling parameter $\left(m\geq\frac{1}{2},\, \Omega_m>0\right)$. 

%====================================================================================================================
% # IV. Performance Analysis
%====================================================================================================================

\section{Performance analysis} \label{performanceAnalysis}
%\subsection{End-to-end SNR statistics}
In this section, we present the performance analysis of the considered system. In subsection~\ref{CDFExpressionForE2ESNR}, a new expression for the CDF of end-to-end SNR is presented. In subsection~\ref{MGFExpressionForE2ESNR}, a new closed-form for the MGF of end-to-end SNR is derived. The new generalized ASER expressions for coherent RQAM and HQAM schemes as well as non-coherent $M$-ary NCFSK scheme, are derived in subsection~\ref{ExacterrorRateExpressions}. In subsection~\ref{AsymptoticerrorRateExpressions}, the asymptotic ASER expression for RQAM scheme and diversity order are given. 
\subsection{CDF expression for end-to-end SNR}  \label{CDFExpressionForE2ESNR}
% ========== Hop 1 and Hop 2 SNR expression ==================
At node R, DF relaying protocol is used, thus, end-to-end SNR, $\lambda_e$, of the mixed THz-RF link can be written as
\begin{align}\label{decode-and-forward}
    \lambda_e = \text{min}(\lambda_1, \lambda_2),
\end{align}
where $\lambda_{1}$ and $\lambda_{2}$ are the instantaneous SNRs of S-R (THz) and R-D (RF) links, respectively, which can be expressed as

\begin{align}\label{SNR1_eqn}
    \lambda_1 = \left|h_{sr}\right|^2 \frac{P_s}{N_1} =\frac{|h_{d1}|^2\,|h_{a1}|^2\,|h_{f1}|^2\,|h_{p1}|^2\,P_s}{N_1},
\end{align}
\begin{align}
   \lambda_2 = \left|h_{rd}\right|^2 \frac{P_r}{N_2} = \frac{|h_{d2}|^2\,|h_{f2}|^2\,P_r}{N_2} \label{SNR2_eqn}.
\end{align}
\newline
We assume that the noise variances are same for both the nodes i.e. $\left(N_1=N_2\right)$. Therefore, we define the average transmit SNR as $\Bar{\lambda}_{0}=\frac{P_{s}}{N_{1}}=\frac{P_{r}}{N_2}$. Further, the CDF expression for $\lambda_{e}$ can be written as

\begin{align} \label{e2e_SNR_eqn}
    F_{\lambda_{e}}(\lambda) = F_{\lambda_{1}}(\lambda) + F_{\lambda_{2}}(\lambda) - F_{\lambda_{1}}(\lambda) F_{\lambda_{2}}(\lambda),
\end{align}
\newline
where $F_{\lambda_{1}}(\lambda)$ and $F_{\lambda_{2}}(\lambda)$ are the CDF expressions of $\lambda_{1}$ and $\lambda_{2}$, respectively and can be obtained with the aid of \cite[eq. (5.4)]{edition2002probability}. The CDF of instantaneous SNR of THz link can be expressed using (\ref{hop1_cdf}) as in (\ref{hop1_cdf_snr}).

\begin{align} \label{hop1_cdf_snr}
    F_{\lambda_{1}}(\lambda) = 1 - \frac{1}{\Gamma\left(\mu\right)}\frac{\phi}{\alpha}G_{2,3}^{3,0}\left[ \mu\left(\mathcal{A}\lambda\right)^{\alpha/2} \left| \begin{array}{c}
    1,\frac{\phi}{\alpha}+1 \\
    \mu,0,\frac{\phi}{\alpha}
    \end{array} \right.\right],
\end{align}
where $\mathcal{A}=\frac{N_1}{P_s\, |h_{d1}|^2\, |h_{a1}|^2\, \Omega^{2}\, S_{0}^{2}}$. The CDF expression of instantaneous SNR of RF link, $\lambda_{2}$, can be expressed as
\newline
\begin{align} \label{hop2_cdf_snr}
    F_{\lambda_2}(\lambda)=1-\frac{\Gamma\left(m, \mathcal{C} \lambda\right)}{\Gamma(m)},
    \end{align}
where $\mathcal{C} = \frac{m N_{2}}{P_{r}\, |h_{d2}|^2\,\Omega_{\mathrm{m}}}$.
Thus, by substituting (\ref{hop1_cdf_snr}) and (\ref{hop2_cdf_snr}) into (\ref{e2e_SNR_eqn}) and after simplification, we can obtain the CDF expression of end-to-end SNR, $\lambda_{e}$, as
% =========== Outage Probability expression without constants ================
\begin{align}\label{e2e_CDF}
%\fontsize{0.25}{0.25}\selectfont
    F_{\lambda_{e}}(\lambda) = 1-\mathcal{B} \, G_{2,3}^{3,0}\left[ \mu\left(\mathcal{A}\lambda\right)^{\alpha/2} \left| \begin{array}{c}
    1,\frac{\phi}{\alpha}+1 \\
    \mu,0,\frac{\phi}{\alpha}
    \end{array} \right.\right] \Gamma \left( m, \mathcal{C}\lambda \right),
\end{align}
where the constant term $\mathcal{B}=\frac{1}{\Gamma\left(\mu\right)\,\Gamma(m)}\frac{\phi}{\alpha}$.

% =========== Moment Generating Function ================
\subsection{MGF expression of end-to-end SNR} \label{MGFExpressionForE2ESNR}
We can evaluate MGF expression of end-to-end SNR, $\lambda_{e}$, with the aid of CDF-based approach as \cite[eq. (6)]{DIXIT2021153883}
\begin{align} \label{MGF_eqn}
   \mathcal{M}_{\lambda_e}(s)=\int_{\lambda=0}^{\infty} s\: e^{-s \lambda} F_{\lambda_{e}}(\lambda) d \lambda,
\end{align}
where $s$ is the Laplace variable. Further, substituting (\ref{e2e_CDF}) into (\ref{MGF_eqn}), the MGF expression can be written as
\begin{align} \label{MGF_eqn_I13}
    \mathcal{M}_{\lambda_e}\left(s\right) = s\mathcal{I}_1(0,s) - s\mathcal{B}\, \mathcal{I}_2(0,s),
\end{align}
where, 
\begin{align} \label{intLambdaE1}
    \mathcal{I}_1 \left(\chi_{1}, \chi_{2} \right) = \int_{\lambda = 0}^{\infty} \lambda^{\chi_{1}} e^{-\chi_{2} \lambda}  d\lambda,
\end{align}
and 
\begin{align} \label{intLambdaEGammaGamma1}
    \mathcal{I}_2 & \left(\chi_1, \chi_2\right) \nonumber \\
    &= \int_{\lambda = 0}^{\infty} \frac{\lambda^{\chi_1} \Gamma\left(m, \mathcal{C} \lambda\right)}{e^{\chi_2 \lambda}\, }  G_{2,3}^{3,0}\left[ \mu\left(\mathcal{A}\lambda\right)^{\alpha/2} \left| \begin{array}{c}
    1,\frac{\phi}{\alpha}+1 \\
    \mu,0,\frac{\phi}{\alpha}
    \end{array} \right.\right] d\lambda.
\end{align}
A solution for both $\mathcal{I}_1(\cdot,\cdot)$ and $\mathcal{I}_2(\cdot,\cdot)$ integrations are given in subsection~\ref{I1sol} and subsection \ref{I2sol} of the appendix, respectively. Substituting the results of $\mathcal{I}_1(\cdot,\cdot)$ and $\mathcal{I}_2(\cdot,\cdot)$ into (\ref{MGF_eqn_I13}) and after simplification, the MGF expression of $\lambda_{e}$ can be expressed as
\begin{align} \label{overall_MGF}
   & \mathcal{M}_{\lambda_e}(s) = 1 - \mathcal{B}\,H_{1,0:2,3;1,2}^{0,1:3,0;2,0}\left[ \left.
    % Column 1 ========
    \begin{array}{c} 
        \mu \left(\sqrt{\frac{\mathcal{A}}{s}}\right)^{\alpha} \\
                                                               \\
        \frac{\mathcal{C}}{s}
    \end{array}
    % Column 2 ========
    \right|
    \begin{array}{c}
    \nu_{\mathcal{M}}:\nu_2;\nu_3 \\ \\
    -:\nu_4;\nu_5
    \end{array}
    \right],
\end{align}
where, $\nu_{\mathcal{M}}$ can be expressed as $\nu_{\mathcal{M}} = \nu_1\left(0\right) = \left\{\left(0;\frac{\alpha}{2},1\right)\right\}$. The values of $\nu_1$ to $\nu_5$ are defined in subsection~\ref{I2sol} of the appendix.

%====================================================================================================================
% # V. Error Rate Expressions
%====================================================================================================================
\subsection{Exact ASER expressions}  \label{ExacterrorRateExpressions}
The well known CDF-based approach can be used to evaluate ASER expression of any modulation scheme in a communication system as \cite[eq. (5)]{8053787}
\begin{align} \label{main_ASER_eqn}
    \mathcal{P}_s(e)=-\int_{\lambda=0}^{\infty} \mathcal{P}_s^{\prime}(e|\lambda) F_{\lambda_e}(\lambda) d\lambda,
\end{align}
where $\mathcal{P}_s^{\prime}(e|\lambda)$ represents the first order derivative of conditional symbol error rate (SER) expression $\mathcal{P}_s(e|\lambda)$ related to the particular modulation scheme and $F_{\lambda_e}(\lambda)$ is the CDF expression for end-to-end SNR, $\lambda_{e}$, of the considered system.
%====================================================================================================================
% Coherent RQAM   #
%====================================================================================================================

\subsubsection{Coherent RQAM scheme}  \label{coherentRQAM}
The first order derivative of conditional SER expression for $M=M_I\times M_Q$-ary coherent RQAM modulation scheme is expressed as \cite[eq. (4)]{9350626}

% =========== Conditional SER expression ================

\begin{align} \label{rqam_cond_SER}
    \mathcal{P}_{s}^{\prime R}(e\mid\lambda)
    =&\frac{1}{\sqrt{\lambda}}\left[\mathcal{D}\ e^{-\left(0.5\,\lambda\,a^2\right)}+\mathcal{F}\ e^{-\left(0.5\,\lambda\,b^2\right)} \right] \nonumber\\
    -&\frac{\mathcal{G}}{\sqrt{\pi}}\ e^{-\frac{\lambda}{2}\left(a^2+b^2\right)}\,{}_1 F_1\left(1;1.5;0.5\,\lambda\, a^2\right) \nonumber\\
    -&\frac{\mathcal{G}}{\sqrt{\pi}}\ e^{-\frac{\lambda}{2}\left(a^2+b^2\right)}\,{}_1 F_1\left(1;1.5;0.5\,\lambda\, b^2\right),
\end{align}
where $ \mathcal{D} = \frac{ap(q-1)}{\sqrt{2\pi}}$, $\mathcal{F} = \frac{b(p-1)q}{\sqrt{2\pi}}$, $\mathcal{G} = \frac{abpq}{\sqrt{\pi}}$, wherein
$p=1-\frac{1}{M_I}$, $q=1-\frac{1}{M_Q}$, $a=\sqrt{\frac{6}{\left(M_{I}^{2}-1\right)+\left(M_{Q}^{2}-1\right) \beta^{2}}}$, $b=\beta a$. $M_I$ is number of in-phase constellation points and $M_Q$ is the number of quadrature-phase constellation points. $\beta$ is the ratio of in-phase and quadrature-phase decision distances. 
Further, the generalized ASER expression of coherent RQAM can be obtained by putting (\ref{e2e_CDF}) and (\ref{rqam_cond_SER}) into (\ref{main_ASER_eqn}) as
\begin{align} \label{rqam_cond_SER_I1234}
    \mathcal{P}_s^R&(e) = -\mathcal{DI}_1\left(-\frac{1}{2},\frac{a^2}{2}\right)-\mathcal{FI}_1\left(-\frac{1}{2},\frac{b^2}{2}\right) \nonumber \\
    &+\frac{\mathcal{G}}{\sqrt{\pi}}\left\{\mathcal{I}_3\left(0,\frac{a^2+b^2}{2}, \frac{a^2}{2}\right)+\mathcal{I}_3\left(0,\frac{a^2+b^2}{2}, \frac{b^2}{2}\right)\right\} \nonumber \\
    &+\mathcal{BD} \, \mathcal{I}_{2}\left(-\frac{1}{2}, \frac{a^2}{2}\right) +\mathcal{BF} \, \mathcal{I}_{2}\left(-\frac{1}{2}, \frac{b^2}{2}\right) \nonumber \\
    &-\frac{\mathcal{BG}}{\sqrt{\pi}}\left\{ \mathcal{I}_{4} \left( \frac{a^2+b^2}{2}, \frac{a^2}{2} \right) + \mathcal{I}_{4} \left( \frac{a^2+b^2}{2}, \frac{b^2}{2} \right) \right\}, 
\end{align}
where, 
\begin{align} \label{intE1F1_1}
    \mathcal{I}_3 \left(\chi_{1}, \chi_{2}, \chi_3 \right) = \int_{\lambda = 0}^{\infty} \frac{\lambda^{\chi_1}}{e^{\chi_{2} \lambda}} { }_{1}{F}_{1}\left(1;\frac{3}{2};\chi_{3} \lambda \right) d\lambda,
\end{align}
and
\begin{align} \label{intLambdaEGammaGamma1F1_1}
    %\mathcal{I}_4 &\left(\chi_1, \chi_2, \chi_3, \chi_4 \right) \nonumber \\
    %&= \mathlarger{\int}_{\lambda = 0}^{\infty} \lambda^{\chi_1} e^{-\chi_2 \lambda} \Gamma(m, \mathcal{C} \lambda) \Gamma\left[\chi_3-\frac{\phi}{\alpha}, \mu(\mathcal{A} \lambda)^{\frac{\alpha}{2}}\right] \nonumber \\
    %& \times { }_{1} F_{1}\left(1 ; \frac{3}{2} ; \chi_{4} \lambda \right)d\lambda.
    \mathcal{I}_4 \left(\chi_1, \chi_2 \right) 
    % \nonumber \\
    &= \int_{\lambda = 0}^{\infty} e^{-\chi_1 \lambda} G_{2,3}^{3,0}\left[ \mu\left(\mathcal{A}\lambda\right)^{\alpha/2} \left| \begin{array}{c}
    1,\frac{\phi}{\alpha}+1 \\
    \mu,0,\frac{\phi}{\alpha}
    \end{array} \right.\right] \nonumber \\
    & \times \Gamma(m, \mathcal{C} \lambda) { }_{1} F_{1}\left(1 ; \frac{3}{2} ; \chi_{2} \lambda \right)d\lambda.
\end{align}
A solution for  $\mathcal{I}_1(\cdot,\cdot)$, $\mathcal{I}_3(\cdot,\cdot,\cdot)$, $\mathcal{I}_2(\cdot,\cdot)$ and $\mathcal{I}_4(\cdot,\cdot)$ are provided in Appendix. With the aid of solutions of $\mathcal{I}_1(\cdot,\cdot)$, $\mathcal{I}_3(\cdot,\cdot,\cdot)$, $\mathcal{I}_2(\cdot,\cdot)$ and $\mathcal{I}_4(\cdot,\cdot)$, the generalized ASER expression for coherent RQAM modulation scheme is presented in (\ref{RQAM_ASER}) in terms of $\Psi_1\left(\cdot\right)$ and $\Psi_2\left(\cdot,\cdot\right)$ function. Where, $\Psi_1\left(\cdot\right)$ and $\Psi_2\left(\cdot,\cdot\right)$ are represented in (\ref{RQAMPsi1}) and (\ref{RQAMPsi2}) respectively.
\begin{align} \label{RQAMPsi1}
    &\Psi_1\left(\chi\right) = H_{1,0:2,3;1,2}^{0,1:3,0;2,0}\left[ \left.
    % Column 1 ========
    \begin{array}{c} 
        \mu \left(\frac{2\mathcal{A}}{\chi}\right)^{\frac{\alpha}{2}} \\
                                                               \\
        \frac{2\mathcal{C}}{\chi}
    \end{array}
    % Column 2 ========
    \right|
    % Column 3 ========
    \begin{array}{c}
    \nu_\mathcal{R}:\nu_2;\nu_3 \\ \\
    -:\nu_4;\nu_5
    \end{array}
    \right], \nonumber \\
\end{align}
where $\nu_\mathcal{R}$ can be expressed as $\nu_\mathcal{R} = \nu_1\left(-\frac{1}{2}\right) = \left\{\left(\frac{1}{2};\frac{\alpha}{2},1\right)\right\}$.
\begin{align} \label{RQAMPsi2}
    &\Psi_2(\chi_1, \chi_2) 
    % = \frac{1}{\chi_2} 
    % \times 
    \nonumber \\
    & = \frac{1}{\chi_2} H_{1,0:1,2;2,3;1,2}^{0,1:1,0;3,0;2,0} \left[ 
    \left.\begin{array}{c} 
        \left(\frac{\chi_1}{\chi_2}\right)  \\ \\
        \mu\left(\frac{2\mathcal{A}}{\chi_2}\right)^{\alpha/2} \\ \\
        \left(\frac{2\mathcal{C}}{\chi_2}\right)
    \end{array}
    \right|
    \begin{array}{c} 
        \nu_6:\nu_7;\nu_2;\nu_3 \\
        -:\nu_8;\nu_4;\nu_5
    \end{array}
    \right],
\end{align}
where, $\nu_6$ to $\nu_8$ are mentioned in subsection~\ref{I4sol} of the appendix. It is well known that SQAM is a special case of RQAM when in-phase and quadrature-phase components are equal to $\sqrt{M}$. Thus, we can obtain an ASER expression for SQAM by putting $M_I=M_Q=\sqrt{M}$ in (\ref{RQAM_ASER}). We can also get the average bit error rate (ABER) expression of BPSK scheme from (\ref{RQAM_ASER}) as a special case of RQAM by putting $M_I=2$, $M_Q=1$, $p=0.5$, $q=0$, $a=\sqrt{2}$ and $\beta=0$.
\begin{figure*}[t!]
\begin{align} \label{RQAM_ASER}
    \mathcal{P}_s^{R}(e) = &\left(p+q-2pq\right) + \frac{2\mathcal{G}}{\sqrt{\pi}}\frac{1}{\left(a^2+b^2\right)} \left[ {}_{2}F_{1}\left(1,1;\frac{3}{2};\frac{a^2}{a^2+b^2}\right)+{}_{2}F_{1}\left(1,1;\frac{3}{2};\frac{b^2}{a^2+b^2}\right)\right] 
% Reference: https://www.overleaf.com/learn/latex/Brackets_and_Parentheses    
    + \frac{\mathcal{B}p\left(q-1\right)}{\sqrt{\pi}} \Psi_1\left(a^2\right) \nonumber \\
    & + \frac{\mathcal{B}\left(p-1\right)q}{\sqrt{\pi}} \Psi_1\left(b^2\right)-\frac{\mathcal{BG}}{\sqrt{\pi}}\Psi_2\left(a^2,b^2\right)-\frac{\mathcal{BG}}{\sqrt{\pi}}\Psi_2\left(b^2,a^2\right)
\end{align}
\hrulefill
\end{figure*}
% ====================================================
%               # V. ASER Analysis/HQAM #
%=====================================================
\subsubsection{Coherent HQAM scheme}  \label{asymptoticHQAM}
The first order derivative of conditional SER of  $M$-ary HQAM scheme can be defined as \cite[eq. (6)]{8053787}
% =========== Conditional SER expression for HQAM modulation Scheme ================
\begin{align}\label{HQAM_cond_SER}
  \mathcal{P}^{\prime H}_{s}(e \mid \lambda) =&  \sqrt{\frac{\alpha_{h}}{2 \pi \lambda}}\left(\frac{B_{c}-B}{2}\right) e^{-\left(\frac{\alpha_{h}}{2}\lambda \right)}
  -\sqrt{\frac{\alpha_{h}}{3 \pi \lambda}}\left(\frac{B_{c}}{3}\right)\nonumber\\
  & \times  e^{-\left(\frac{\alpha_{h}}{3}\lambda \right)} +\sqrt{\frac{\alpha_{h}}{6 \pi \lambda}}\left(\frac{B_{c}}{2}\right)  e^{-\left(\frac{\alpha_{h} }{6}\lambda \right)}\nonumber\\
  &+\frac{2 B_{c} \alpha_{h}}{9 \pi} { }_{1} F_{1}\left(1 ; \frac{3}{2} ; \frac{\alpha_{h}}{3}\lambda \right)e^{-\left(\frac{2 \alpha_{h} }{3} \lambda \right)}\nonumber\\
  &-\frac{B_{c} \alpha_{h}}{2 \sqrt{3} \pi}\left[ { }_{1} F_{1}\left(1 ; \frac{3}{2} ; \frac{\alpha_{h}}{2}\lambda \right)\right.\nonumber\\
  &+\left.{ }_{1} F_{1}\left(1 ; \frac{3}{2} ; \frac{\alpha_{h}}{6}\lambda \right)\right]e^{-\left(\frac{2 \alpha_{h}}{3}\lambda \right)},
\end{align}
where $B$, $B_c$, $\alpha_h$ are different modulation parameters as defined in \cite[Table-I]{9079579} in terms of $\tau$, $\tau_c$ and $K$, respectively. Similar to RQAM, the ASER for HQAM scheme can be evaluated by putting (\ref{e2e_CDF}) and (\ref{HQAM_cond_SER}) into (\ref{main_ASER_eqn}) as
\begin{align}\label{HQAM_cond_SER_I1234}
    &\mathcal{P}_s^H(e) = \left( \frac{B-B_c}{2}\right)\sqrt{\frac{\alpha_h}{2\pi}} \mathcal{I}_1\left(-\frac{1}{2},\frac{\alpha_h}{2} \right) \nonumber \\
    & +\frac{B_c}{3}\sqrt{\frac{\alpha_h}{3\pi}}\mathcal{I}_1\left(-\frac{1}{2}, \frac{\alpha_h}{3}\right) - \frac{B_c}{2}\sqrt{\frac{\alpha_h}{6\pi}}\mathcal{I}_1\left(-\frac{1}{2},\frac{\alpha_h}{6}\right) \nonumber \\
    & + \frac{B_c \alpha_h}{2\sqrt{3} \pi} \left\{\mathcal{I}_3\left(0,\frac{2\alpha_h}{3}, \frac{\alpha_h}{2}\right)+\mathcal{I}_3\left(0,\frac{2 \alpha_h}{3}, \frac{\alpha_h}{6}\right)\right \} \nonumber \\
    & - \frac{2B_{c} \alpha_h}{9\pi} \mathcal{I}_3\left(0,\frac{2 \alpha_h}{3}, \frac{\alpha_h}{3}\right) 
    % \nonumber \\
    % & 
    + \mathcal{B}  \left[ \sqrt{\frac{\alpha_h}{2\pi}}\left(\frac{B_c-B}{2}\right) \right. 
    \nonumber \\
    & 
   \times  \mathcal{I}_2\left(-\frac{1}{2}, \frac{\alpha_h}{2}\right) 
    % \nonumber \\
    % & 
    - \sqrt{\frac{\alpha_h}{3\pi}}\left(\frac{B_c}{3}\right) \mathcal{I}_2\left(-\frac{1}{2},\frac{\alpha_h}{3}\right) \nonumber \\
    & + \sqrt{\frac{\alpha_h}{6\pi}}\left(\frac{B_c}{2}\right) \mathcal{I}_2\left(-\frac{1}{2},\frac{\alpha_h}{6}\right) 
    % \nonumber \\
    % & 
    + \frac{2B_{c} \alpha_h}{9\pi} \mathcal{I}_4 \left(\frac{2 \alpha_h}{3},\frac{\alpha_h}{3}\right) \nonumber \\
    & - \frac{B_c \alpha_h}{2\sqrt{3}\pi} \left. \left\{\mathcal{I}_4\left(\frac{2 \alpha_h}{3},\frac{\alpha_h}{2}\right) + \mathcal{I}_4\left(\frac{2 \alpha_h}{3},\frac{\alpha_h}{6}\right) \right\} \right].
\end{align}
Note that a solution for $\mathcal{I}_1(\cdot,\cdot)$, $\mathcal{I}_3(\cdot,\cdot,\cdot)$, $\mathcal{I}_2(\cdot,\cdot)$ and $\mathcal{I}_4(\cdot,\cdot)$ are provided in Appendix. Using the solutions of $\mathcal{I}_1(\cdot,\cdot)$, $\mathcal{I}_3(\cdot,\cdot,\cdot)$, $\mathcal{I}_2(\cdot,\cdot)$ and $\mathcal{I}_4(\cdot,\cdot)$, the ASER expression of generalized HQAM scheme can be expressed in terms of $\Psi_3\left(\cdot\right)$ and $\Psi_4\left(\cdot,\cdot\right)$ functions as represented in (\ref{HQAM_ASER}). Where the functions $\Psi_3\left(\cdot\right)$ and $\Psi_4\left(\cdot,\cdot\right)$ are represented in (\ref{HQAMPsi3}) and (\ref{HQAMPsi4}) respectively.
\begin{align} \label{HQAMPsi3}
    &\Psi_3\left(\chi\right) = H_{1,0:2,3;1,2}^{0,1:3,0;2,0}\left[ \left.
    % Column 1 ========
    \begin{array}{c} 
        \mu \left(\frac{\mathcal{A}\chi}{\alpha_h}\right)^{\frac{\alpha}{2}} \\
                                                               \\
        \frac{\mathcal{C}\chi}{\alpha_h}
    \end{array}
    % Column 2 ========
    \right|
    % Column 3 ========
    \begin{array}{c}
    \nu_\mathcal{H}:\nu_2;\nu_3 \\ \\
    -:\nu_4;\nu_5
    \end{array}
    \right], \nonumber \\
\end{align}
where $\nu_\mathcal{H}$ can be expressed as $\nu_\mathcal{H} = \nu_1\left(-\frac{1}{2}\right) = \left\{\left(\frac{1}{2};\frac{\alpha}{2},1\right)\right\}$.
\begin{align} \label{HQAMPsi4}
  &  \Psi_4(\chi_1, \chi_2) \nonumber \\
    &=H_{1,0:1,2;2,3;1,2}^{0,1:1,0;3,0;2,0} \left[ 
    \left.\begin{array}{c} 
        \chi_1 \\ \\
        \mu\left(\frac{\mathcal{A}\chi}{\alpha_h}\right)^{\alpha/2} \\ \\
        \left(\frac{\mathcal{C}\chi_2}{\alpha_h}\right)
    \end{array}
    \right|
    \begin{array}{c} 
        \nu_6:\nu_7;\nu_2;\nu_3 \\
        -:\nu_8;\nu_4;\nu_5
    \end{array}
    \right].
\end{align}
% =========== HQAM ASER Expression ================
\begin{figure*}[t!]
\begin{align} \label{HQAM_ASER}
    \mathcal{P}_s^{H}(e) = & \frac{B}{2} - \frac{B_c}{3} - \frac{\mathcal{B}}{\sqrt{\pi}}\left\{\left(\frac{B_c-B}{2}\right)\Psi_3\left(2\right)-\frac{B_c}{3}\Psi_3\left(3\right)+\frac{B_c}{2}\Psi_3\left(6\right)\right\} + \frac{\mathcal{B}B_c}{\pi}\left\{\frac{\Psi_4\left(1,3\right)}{3} -\frac{\sqrt{3}\Psi_4\left(3,6\right)}{2}-\frac{\Psi_4\left(\frac{1}{3},2\right)}{2\sqrt{3}}\right\}
\end{align}
\hrulefill
\end{figure*}
\newline

%====================================================================================================================
% # V. ASER Analysis/NCFSK #
%====================================================================================================================
\subsubsection{$M$-ary NCFSK scheme}  \label{NCFSK}
For any modulation scheme, the ASER expression can be evaluated by using the PDF-based approach \cite{4544952} as
\begin{align}\label{Aser_From_PDF}
    \mathcal{P}_s(e)=\int_{\lambda=0}^\infty \mathcal{P}_s(e|\lambda) f_{\lambda_{e}}(\lambda) d\lambda,
\end{align}
where $f_{\lambda_e}(\lambda)$ is the PDF expression for end-to-end SNR, $\lambda_{e}$, of the considered system. The conditional SER for $M$-ary NCFSK modulation scheme  can be written \cite[eq. (8.67)]{4544952} as
\begin{align}\label{ncfsk_conditional_ser}
    \mathcal{P}_{s}^{NC}(e|\lambda)=\sum_{\eta=1}^{M-1} \frac{(-1)^{\eta+1}}{\eta+1} \binom{ M-1}{\eta} e^{-\left(\frac{\eta\lambda}{\eta+1}\right)}.
\end{align}
By putting (\ref{ncfsk_conditional_ser}) into (\ref{Aser_From_PDF}), the ASER expression of $M$-ary NCFSK scheme can be expressed as
\begin{align}\label{ASER_NCFSK_intermediate}
    P_{e}^{NC}=\sum_{\eta=1}^{M-1} \frac{(-1)^{\eta+1}}{\eta+1}\left(\begin{array}{c} M-1 \\ \eta \end{array}\right) \int_{\lambda=0}^\infty e^{-\left(\frac{\eta\lambda}{\eta+1}\right)} f_{\lambda_{e}}(\lambda) d\lambda.
\end{align}
Further, with the aid of PDF based approach of MGF calculation \cite[eq. (2.4)]{4544952}, the ASER expression of $M$-ary NCFSK scheme can be evaluated in terms of the MGF as 
\begin{align}\label{ASER_NCFSK_in_term_of_MGF}
    P_{e}^{NC}=\sum_{\eta=1}^{M-1} \frac{(-1)^{\eta+1}}{\eta+1}\left(\begin{array}{c} M-1 \\ \eta \end{array}\right) \mathcal{M}_{\lambda_e}\left(\frac{\eta}{\eta+1}\right).
\end{align}
Further, by substituting (\ref{overall_MGF}) into (\ref{ASER_NCFSK_in_term_of_MGF}) and putting $M=2$, we can write the simplified ABER expression for binary NCFSK as 
\begin{align}
    \mathcal{P}_e^{NC} = \frac{1}{2} - \frac{\mathcal{B}}{2} H_{1,0:2,3;1,2}^{0,1:3,0;2,0}\left[ \left.
    % Column 1 ========
    \begin{array}{c} 
        \mu \left(\sqrt{2\mathcal{A}}\right)^{\alpha} \\
                                                      \\
        2\mathcal{C}
    \end{array}
    % Column 2 ========
    \right|
    % Column 3 ========
    \begin{array}{c}
    \nu_{\mathcal{M}}:\nu_2;\nu_3 \\ \\
    -:\nu_4;\nu_5
    \end{array}
    \right].
\end{align}
\subsection{Asymptotic ASER Expression and Diversity Order}
\label{AsymptoticerrorRateExpressions}

In order to observe the behavior of ASER in high SNR regime and to examine the system's diversity, we need to derive the asymptotic ASER for the RQAM scheme. To do so, first, we have to obtain the asymptotic expression of (\ref{hop1_cdf_snr}) and (\ref{hop2_cdf_snr}) by considering $\Bar{\lambda}_{0}\rightarrow \infty$ or $P_{s}\rightarrow \infty$. In this context, we approximate the Meijer-G function of (\ref{hop1_cdf_snr}) with the aid of \cite[eq. (07.34.06.0001.01)]{wolfram07.34.06.0001.01} as represented in (\ref{hop1Asymptotic}). 
\begin{align}
&F_{\lambda_1}^{\infty}(\lambda) \approx \frac{\Gamma\left(\mu-\frac{\phi}{\alpha}\right)}{\Gamma(\mu)} \mu^{\frac{\phi}{\alpha}}(\mathcal{A} \lambda)^{\frac{\phi}{2}}+\frac{\phi \mu^\mu(\mathcal{A} \lambda)^{\frac{\alpha \mu}{2}}}{(\phi-\alpha \mu) \Gamma(\mu+1)} \label{hop1Asymptotic}.
\end{align}
Further, we substitute the upper incomplete gamma function by its approximation as $\lim\limits_{x\rightarrow0} \Gamma(a, x) \approx \Gamma(a)-\frac{x^{a}}{a}$ \cite[eq. (8.354.2)]{edition2002probability} to express the asymptotic form of (\ref{hop2_cdf_snr}) as mentioned in (\ref{hop2Asymptotic}).
\begin{align}
F_{\lambda_2}^{\infty}(\lambda) &\approx \frac{(\mathcal{C} \lambda)^{\mathrm{m}}}{\Gamma(\mathrm{m}+1)} \label{hop2Asymptotic}.
\end{align}
Furthermore, the asymptotic CDF expression of $\lambda_e$ can be approximated for DF relaying scheme as 
$F_{\lambda_e}^{\infty}(\lambda) \approx F_{\lambda_1}^{\infty}(\lambda) + F_{\lambda_2}^{\infty}(\lambda)$. $F_{\lambda_e}^{\infty}(\lambda)$ can be obtained by adding (\ref{hop1Asymptotic}) and (\ref{hop2Asymptotic}) after putting the values of $\mathcal{A}$ and $\mathcal{C}$, respectively, as expressed in (\ref{Asymp_outage}), given at the next page.
\newline

\begin{figure*}[t!]
\begin{align} \label{Asymp_outage}
    F_{\lambda_e}^{\infty}\left(\lambda\right) &\approx \left[\frac{\phi \mu^\mu}{(\phi-\alpha \mu) \Gamma(\mu+1)}\left(\frac{\lambda}{\Omega^2 S_0^2\left|h_{d1}\right|^2\left|h_{a1}\right|^{2}}\right)^{\frac{\alpha\mu}{2}}\right]\left(\frac{N_1}{P_s}\right)^{\frac{\alpha\mu}{2}}+\left[\frac{\mu^{\left(\frac{\phi}{\alpha}\right)} \Gamma\left(\mu-\frac{\phi}{\alpha}\right)}{\Gamma(\mu)}\left(\frac{\lambda}{\Omega^{2}S_{0}^{2}\left|h_{d1}\right|^{2}\left|h_{a1}\right|^{2}}\right)^{\frac{\phi}{2}}\right]\left(\frac{N_{1}}{P_{s}}\right)^{\frac{\phi}{2}}\nonumber\\& 
    +\left[\frac{1}{\Gamma(m+1)}\left(\frac{m\lambda}{\Omega_m\left|h_{d2}\right|^2}\right)^m\right]\left(\frac{N_2}{P_r}\right)^m.
\end{align}
\hrulefill
\end{figure*}

%%%========Asymptotic ASER for RQAM scheme===============
\subsubsection{Asymptotic ASER for RQAM scheme} \label{asymptoticRQAm}
Asymptotic ASER expression of RQAM scheme (\ref{RQAM_Asymptotic}) can be evaluated by putting the approximated CDF of end-to-end SNR (\ref{Asymp_outage}) and (\ref{rqam_cond_SER_I1234}) into (\ref{main_ASER_eqn}) as
\begin{align} \label{RQAM_Asymptotic_I}
    \mathcal{P}&_s^{R^\infty}(e) = - \mathcal{DRI}_1\left(\frac{\alpha\mu-1}{2},\frac{a^2}{2}\right)-\mathcal{DTI}_1\left(\frac{\phi-1}{2},\frac{a^2}{2}\right) \nonumber \\
    &-\frac{\mathcal{C}^m}{\Gamma\left(m+1\right)}\left\{\mathcal{DI}_1\left(m-\frac{1}{2},\frac{a^2}{2}\right)+\mathcal{FI}_1\left(m-\frac{1}{2},\frac{b^2}{2} \right)\right\} \nonumber \\
    &-\mathcal{FRI}_1\left(\frac{\alpha\mu-1}{2},\frac{b^2}{2}\right)-\mathcal{FTI}_1\left(\frac{\phi-1}{2},\frac{b^2}{2}\right) \nonumber \\
    &+\frac{\mathcal{G}}{\sqrt{\pi}}\left[\mathcal{RI}_3\left(\frac{\alpha\mu}{2},\frac{a^2+b^2}{2},\frac{a^2}{2}\right)\right. +\mathcal{TI}_3\left(\frac{\phi}{2},\frac{a^2+b^2}{2},\frac{a^2}{2}\right) \nonumber \\
    &+\mathcal{RI}_3\left(\frac{\alpha\mu}{2},\frac{a^2+b^2}{2},\frac{b^2}{2}\right) +\mathcal{TI}_3\left(\frac{\phi}{2},\frac{a^2+b^2}{2},\frac{b^2}{2}\right) \nonumber \\
    &+\frac{\mathcal{C}^m}{\Gamma\left(m+1\right)}\left\{\mathcal{I}_3\left(m,\frac{a^2+b^2}{2},\frac{a^2}{2}\right) \right. \nonumber \\
    &+\left.\left.\mathcal{I}_3\left(m,\frac{a^2+b^2}{2},\frac{b^2}{2}\right)\right\}\right],
\end{align}
where $\mathcal{R}=\frac{\phi\mu^\mu}{\left(\phi-\alpha\mu\right)\Gamma\left(\mu+1\right)}\mathcal{A}^{\frac{\alpha\mu}{2}}, \mathcal{T}=\frac{\mu^{\left(\phi/\alpha\right)}}{\Gamma\left(\mu\right)}\Gamma\left(\mu-\frac{\phi}{\alpha}\right)\mathcal{A}^{\frac{\phi}{2}}$. Now, by substituting the integral values in (\ref{RQAM_Asymptotic_I}), asymptotic ASER expression of RQAM scheme can be written in terms of $\Psi^\infty\left(\cdot,\cdot\right)$ as represented in (\ref{RQAM_Asymptotic}), where $\Psi^\infty\left(\cdot,\cdot\right)$ can be written as $\Psi^\infty\left(\chi_{1},\chi_{2}\right) = \left(\frac{\chi_2}{2}\right)^{-\frac{\chi_1+1}{2}} \Gamma\left(\frac{\chi_1+1}{2}\right)$. \newline
\begin{figure*}[t!]
\begin{align} \label{RQAM_Asymptotic}
     \mathcal{P}&_s^{R^{\infty}}(e) = -\mathcal{R}\bigg[\mathcal{D}\Psi^\infty\left(\alpha\mu,a^2\right)-\mathcal{F}\Psi^\infty\left(\alpha\mu,b^2\right)\bigg]+\frac{\mathcal{G}}{\sqrt{\pi}} \left[{}_{2}F_{1}\left(1,1;\frac{3}{2};\frac{a^2}{a^2+b^2}\right)+{}_{2}F_{1}\left(1,1;\frac{3}{2};\frac{b^2}{a^2+b^2}\right)\right] \times \nonumber \\
     & \left[\mathcal{R}\Psi^\infty\left(\alpha\mu+1,a^2+b^2\right)+\mathcal{T}\Psi^\infty\left(\phi+1,a^2+b^2\right)+\frac{1}{\mathcal{C}}\left(\frac{2\mathcal{C}}{a^2+b^2}\right)^{m+1}\right] -\frac{\mathcal{C}^m}{\Gamma\left(m+1\right)}\bigg[\mathcal{D}\Psi^\infty\left(2m,a^2\right)+\mathcal{F}\Psi^\infty\left(2m,b^2\right)\bigg] \nonumber \\
     &-\mathcal{T}\bigg[\mathcal{D}\Psi^\infty\left(\phi,a^2\right)-\mathcal{F}\Psi^\infty\left(\phi,b^2\right)\bigg].
\end{align}
\hrulefill
\end{figure*}
\subsubsection{Diversity Order} \label{DiversityOrderFromOutage}
Mathematically, the diversity order can be evaluated in high SNR regime as \cite{wang2003simple}
\begin{align}\label{diversity_order_eqn}
   \text{Diversity\ Order}& =\lim _{\Bar{\lambda}_{0} \rightarrow \infty} \frac{-\log P_{s}^{\infty}\left(e\right)}{\log \Bar{\lambda}_{0}}.
\end{align}
Thus, one can obtain the diversity order of the proposed system by putting the values of $\mathcal{A}$ and $\mathcal{C}$ in (\ref{RQAM_Asymptotic}) and substituting resulting expression into (\ref{diversity_order_eqn}), which can be expressed as
\begin{equation}\label{diversity_order_from_rqam}
    \text{Diversity\ Order} = \min\left\{\frac{\alpha \mu}{2}, \frac{\phi}{2}, m\right\}.
\end{equation}
%%% ========Asymptotic ASER for HQAM scheme===============

%====================================================================================================================
% # VI. NUMERICAL AND SIMULATION RESULT #
%====================================================================================================================
\section{Numerical and simulation results} \label{numericalAndSimulationResults}
In this section, we analyze the analytical, simulation and asymptotic results to understand the performance of relay-assisted THz-RF systems. We choose $10^8$ realizations of channel coefficients and symbols in our Monte Carlo simulations to validate the theoretically generated results whereas asymptotic behavior is also observed. We consider environmental conditions such as atmospheric pressure, absolute temperature, and relative humidity for THz link as well as different channel parameters as mentioned in Table \ref{parameter table}.

\begin{table}[h!]
\centering
\caption{\label{parameter table}list of Simulation Parameters}
\resizebox{\columnwidth}{!}{%
     \begin{tabular}{| l | l |}
     \hline
     \centering
     \textbf{Parameter} & \textbf{Value} \\
     \hline
     THz carrier frequency ($f_{sr}$) & 275 GHz \\ \hline
     RF carrier frequency ($f_{rd}$) & 8 GHz \\ \hline
     THz link Antenna Gain ($G_{ts},G_{rr}$) & 52 dBi \\ \hline
     RF link Antenna Gain ($G_{tr},G_{rd}$) & 52 dBi \\ \hline
     Source to Relay distance ($d_{sr}$)& 300 Meters \\ \hline
     Relay to Destination distance ($d_{rd}$) & 800 Meters \\ \hline
     Absolute Temperature ($T$) & 296 K \\ \hline
     Atmospheric Pressure ($P$) & 1013.25 hPa\\ \hline
     Relative Humidty ($\psi$) & 50 \% \\ \hline
     THz link fading parameter ($\alpha$) & 1.6-3.0 \\ \hline
     %THz link Normalised Variance ($\mu$) & 1-3 \\ \hline
     Normalised variance of THz link envelope ($\mu$) & 1-3.5 \\ \hline
     $\alpha$-root mean value of THz link envelope ($\Omega$) & 1-2.5 \\ \hline
     Pointing Error Parameter ($\phi$) & 1.5-12 \\ \hline
     Fraction of power collected at center ($S_0$) & 0.39-0.73 \\ \hline
     RF link Shape Parameter ($m$) & 1.6-3.0 \\ \hline
     RF link Spread Parameter ($\Omega_m$) & 0.015-3.00 \\ \hline 
\end{tabular}
}
\end{table}

\begin{figure}[hb!]
\centering
  \includegraphics[width=\linewidth]{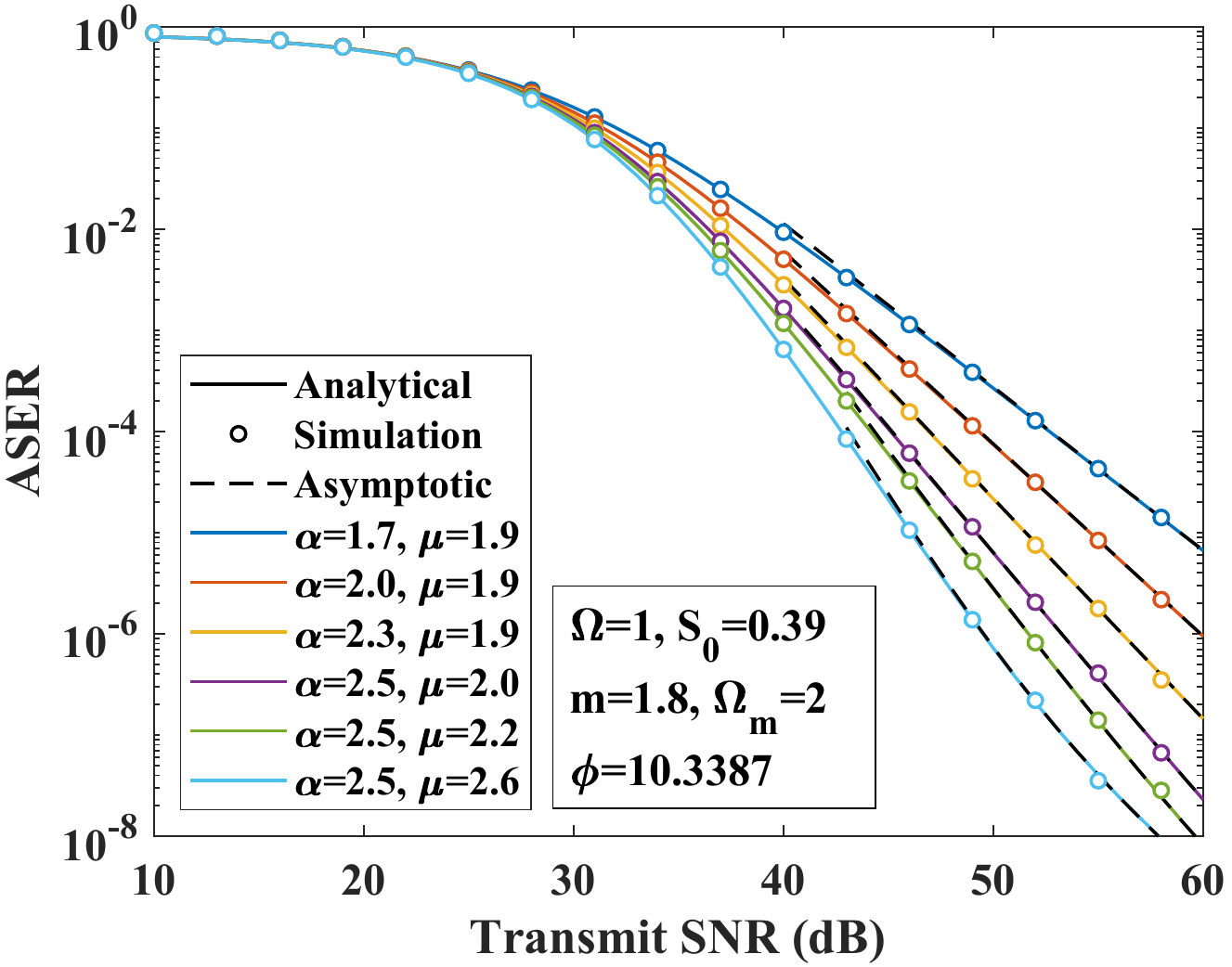}\\
  \caption{ASER performance of $4\times 2$-RQAM scheme versus average SNR. }\label{graph_rqam_alpha_mu}
\end{figure}

Fig. \ref{graph_rqam_alpha_mu} illustrates the ASER performance of $4\times 2$-RQAM scheme versus average transmit SNR for several sets of $\alpha$ and $\mu$. From the figure, we find that the theoretical and simulation results match well to each other which validates the correctness of the derived expression (\ref{RQAM_ASER}). The asymptotic results also align with the theoretical and the simulation results at high SNR regime by which we can further validate the correctness of the derived expression. It is also observed that by increasing $\alpha$ and $\mu$, the ASER performance improves as expected. For example, approximately 4.46 dB and 7.65 dB less SNR is required to maintain the ASER at $6.66\times10^{-6}$ when $\alpha$ is increased from 1.7 to 2 and 1.7 to 2.3, respectively, keeping other parameters constant. Similarly, when $\mu$ is increased from 2.0 to 2.2 and 2.0 to 2.6, approximately, 1.88 dB and 3.93 dB less SNR is required, respectively, to maintain the ASER at $3\times10^{-8}$.

%----------------------------------------Updated------------------------------------
% 4x2 RQAM graph Variation of alpha-mu

\begin{figure}[hb!]
\centering
  \includegraphics[width=\linewidth]{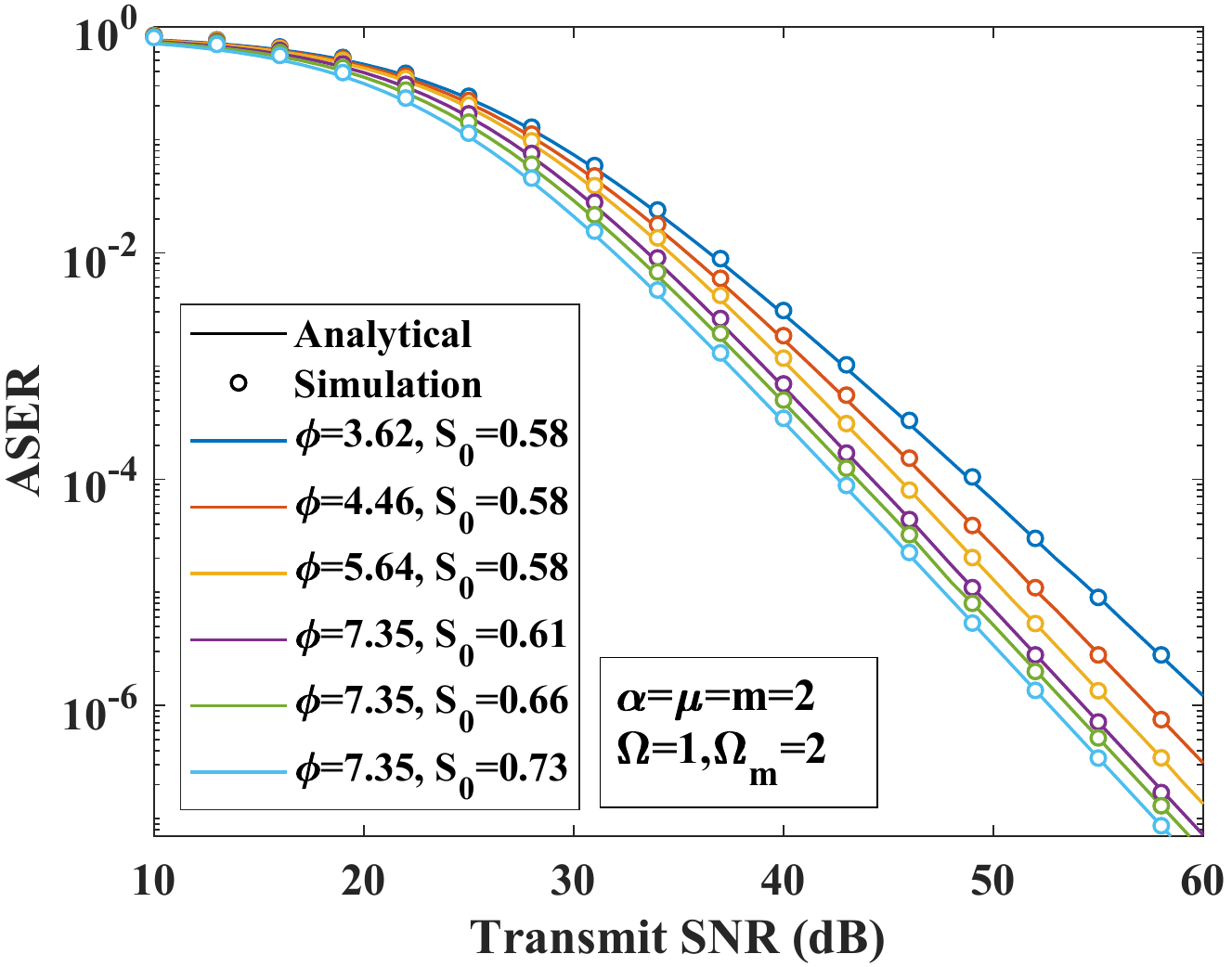}\\
  \caption{ASER performance of $8$-HQAM scheme against average SNR.}\label{graph_hqam_phi_s0}
\end{figure}

In Fig. \ref{graph_hqam_phi_s0}, the ASER performance is analyzed for $8$-HQAM scheme against average transmit SNR in order to demonstrate the impact of pointing error coefficients. Further, it can be observed that the theoretical results match well with the simulation results for all the investigated cases. From the figure, we perceive that the ASER performance is improved with an increase in both the parameters $\phi$ and $S_0$. For example, when $\phi$ is increased from 3.62 to 4.46 and 3.62 to 5.64, approximately 3.05 dB and 4.75 dB less SNR is required, respectively, to achieve the ASER of $1.3\times10^{-6}$. Similarly, when $S_0$ is increased from 0.61 to 0.66 and 0.61 to 0.73, approximately, 0.71 dB and 1.6 dB less SNR is required, respectively, to maintain the ASER of $10^{-7}$ keeping all other parameters constant.

% 8-HQAM graph variation of Phi-S0
%--------------------------------------------------------------------------

\begin{figure}[h!]
\centering
  \includegraphics[width=\linewidth]{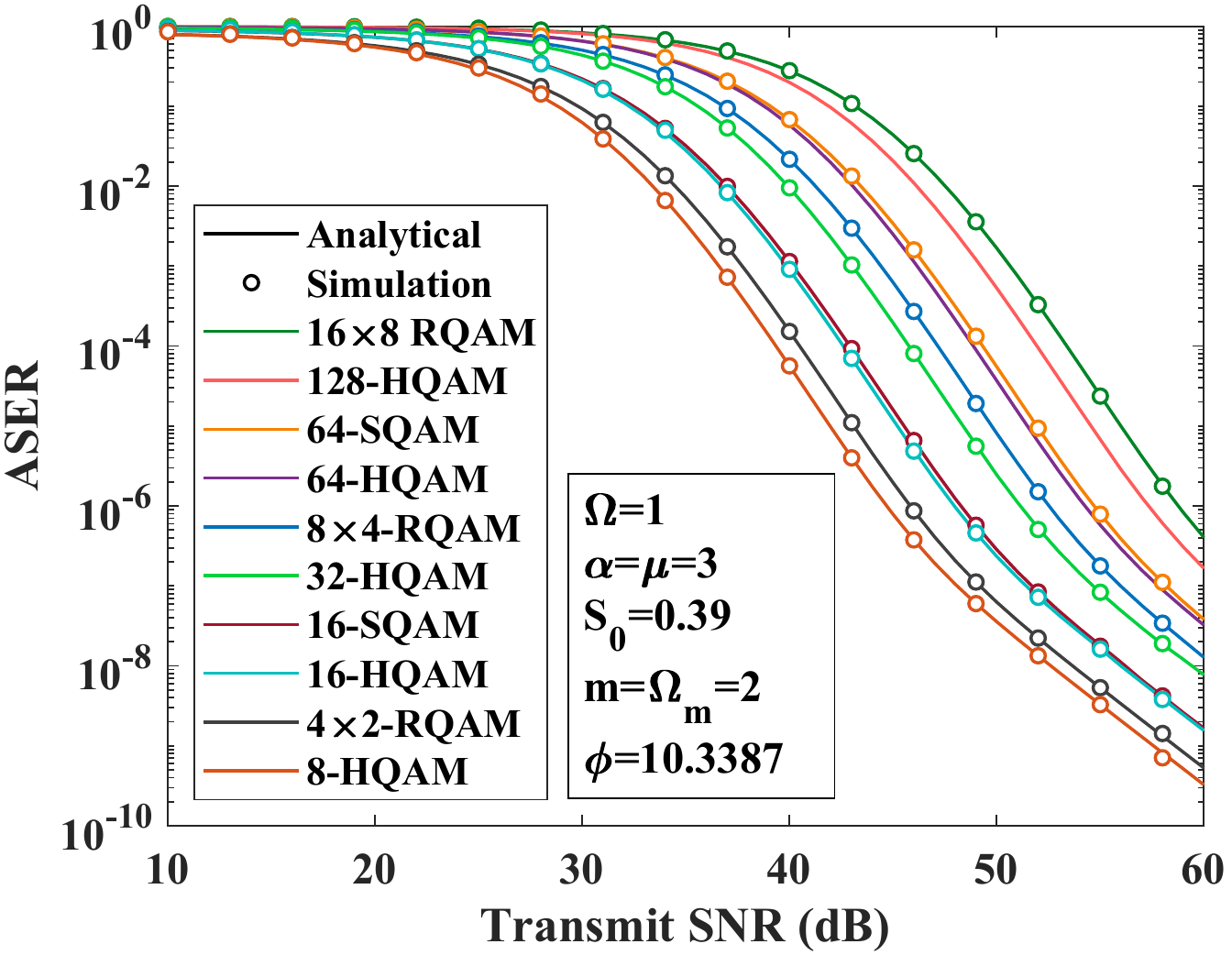}\\
  \caption{Comparison of ASER performances against average SNR for several QAM constellations.}\label{graph_hqam_vs_rqam}
\end{figure}

In Fig. \ref{graph_hqam_vs_rqam}, the ASER performance for several modulation schemes are compared. We observe that, due to the compact packing of constellation points, HQAM has the best performance among all QAM schemes. SQAM is a compact arrangement of RQAM and thus SQAM provides significantly better performance than RQAM. It is noticeable that $64$-HQAM and $16$-HQAM show better ASER performance than $64$-SQAM and $16$-SQAM respectively, though the difference in performance is very less because of their compact constellations. For example, when the modulation scheme is changed from $8\times4$-RQAM to $32$-HQAM, approximately, 1.33 dB less SNR is required to maintain the ASER at $10^{-7}$. Also, ASER performance gets better when modulation order is reduced. From the figure, we can further observe that the higher-order modulation schemes suit perfectly for good channel conditions with high SNR at the receiver, and therefore we get the highest possible data rates without degrading the error performance much. Although, for a low SNR regime, it is better to use lower-order modulation schemes to combat severe degradation in ASER. 
% Comparison between RQAM and HQAM
%----------------------------------------------------------------------------

\begin{figure}[ht!]
\centering
  \includegraphics[width=\linewidth]{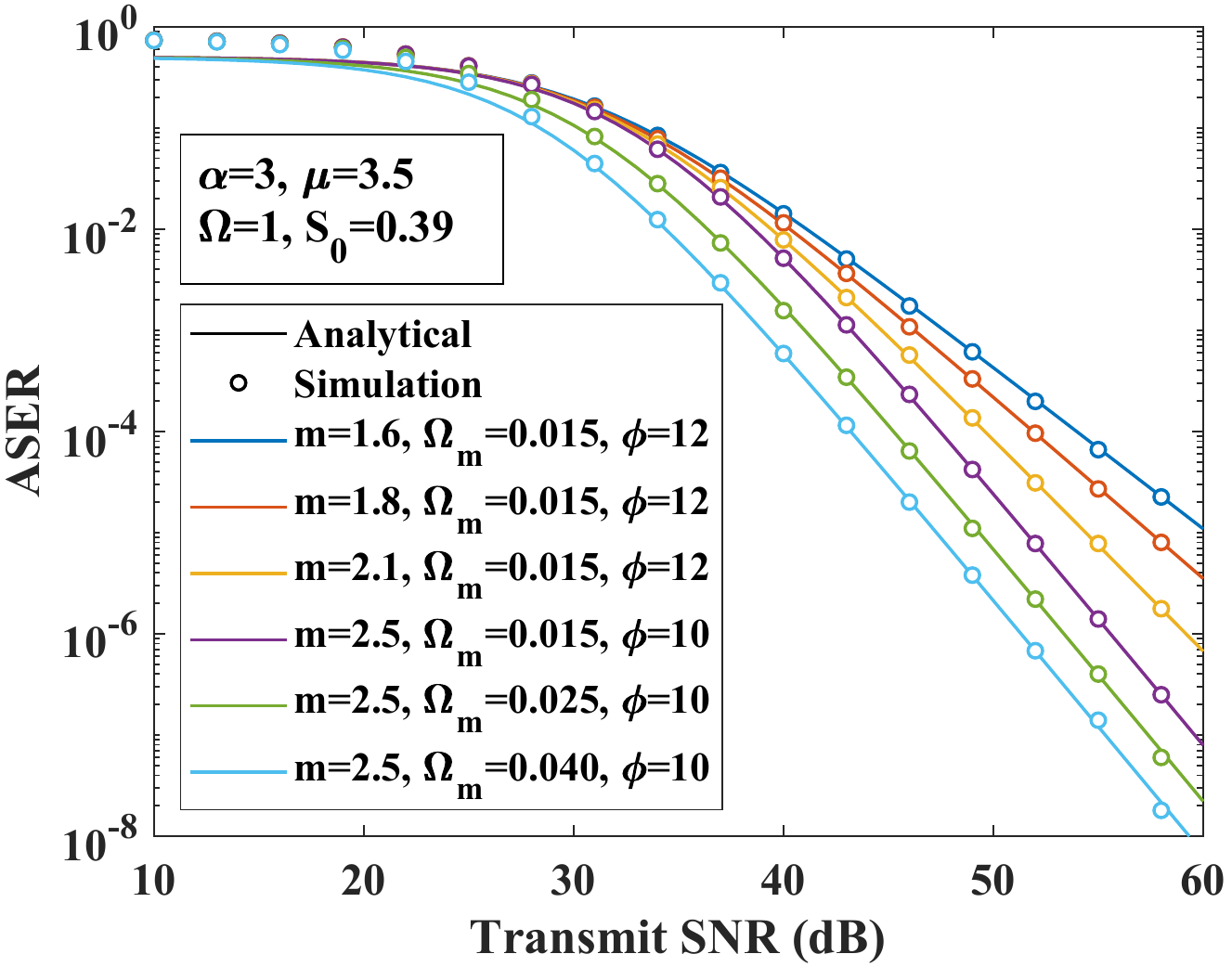}\\
  \caption{ASER performance of $2$-ary NCFSK modulation scheme against average SNR. }\label{graph_ncfsk_m_wm}
\end{figure}

Fig. \ref{graph_ncfsk_m_wm} shows the theoretical and the simulation results of ASER performance with respect to average transmit SNR for $2$-ary NCFSK modulation scheme considering several values of RF link parameters, $m$, and $\Omega_m$, with fixed THz link parameters. From the figure, it can be seen that all the theoretical curves match well with the simulation curves for all the investigated cases. Further, it is also observed that with the increase in $m$ and $\Omega_m$, the ASER performance improves. For example, when $m$ is increased from 1.6 to 1.8 and 1.6 to 2.1, approximately, 2.73 dB and 5.77 dB less SNR is required, respectively, to achieve the ASER of $1.1\times10^{-5}$ keeping all the other parameters constant. Similarly, when $\Omega_m$ is increased from 0.015 to 0.025 and 0.015 to 0.04, approximately, 2.23 dB and 4.27 dB of less SNR is required to maintain the ASER at $10^{-7}$.

% NCFSK graph variation between m and $\omega_m$
\begin{figure}[ht!]
\centering
  \includegraphics[width=\linewidth]{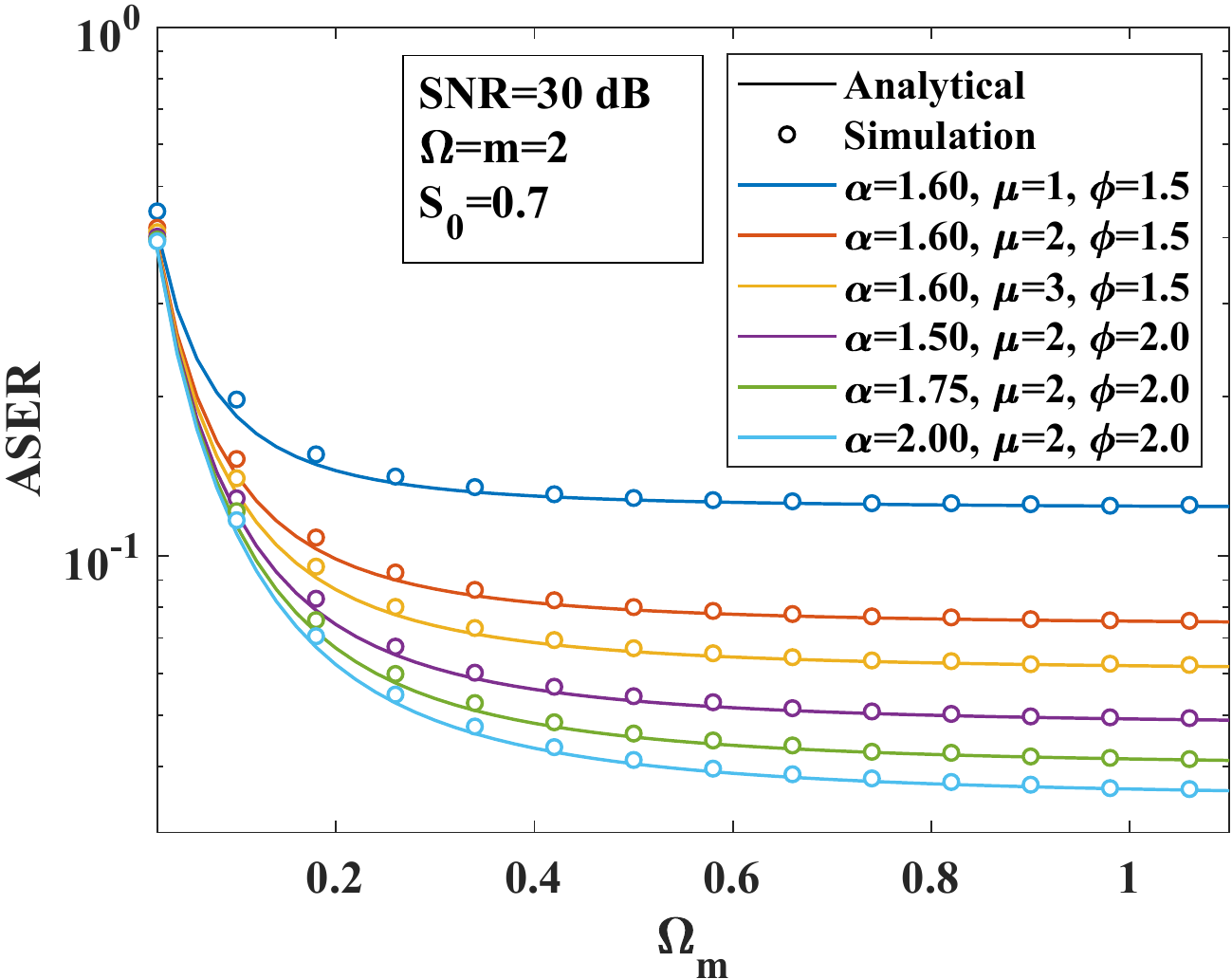}\\
  \caption{ASER performance of $4\times 2$-RQAM scheme against $\Omega_{m}$.}\label{graph_aser_wm}
\end{figure}

%-----------------------------------------------------------------------------
In Fig. \ref{graph_aser_wm}, we show the impact of $\Omega_{m}$ on ASER performance of $4 \times 2$-RQAM schemes, considering several values of $\alpha$, $\mu$ and $\phi$ at fixed average transmit SNR and other parameters. From the figure, it can be observed that all the theoretical and simulation curves match well for all the investigated cases. It is also noticed that with the increase in $\Omega_{m}$ the ASER performance improves and gets saturated further.

% Effect of Phi on ASER
\begin{figure}[ht]
\centering
  \includegraphics[width=\linewidth]{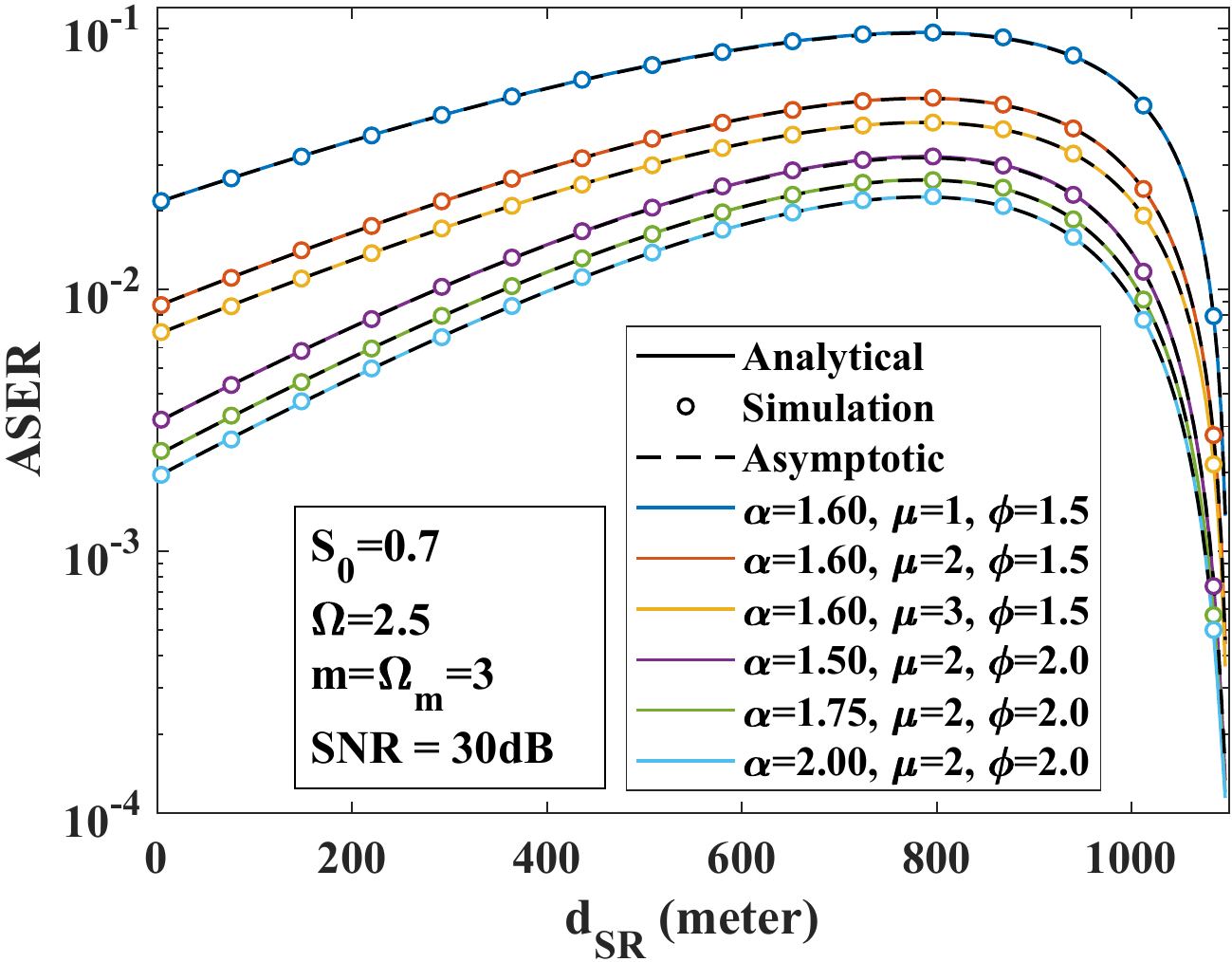}\\
  \caption{ASER performance of $4\times 2$-RQAM scheme against $d_{SR}$.}\label{graph_aser_distance}
\end{figure}

%---------------------------------------------------------------------------
Fig. \ref{graph_aser_distance} represents the impact of $d_{SR}$ on ASER performance of $4\times 2$-RQAM scheme considering several values of $\alpha$ and $\mu$ at fixed average transmit SNR. We consider $d_{SR}+d_{RD}=1100$ (meters). At fixed $d_{SR}$, it is computed as $d_{RD}=1100-d_{SR}$. Further, we can observe that by increasing $d_{SR}$,  the probability of ASER increases, gets its maximum value and then decreases. 
From this observation, it can be depicted that the ASER performance improves when 
relay is near to source or destination node.

% Effect of Distance on ASER (4*2 RQAM)

%====================================================================================================================
% # VII. CONCLUSION #
%====================================================================================================================

\section{Conclusion}  \label{conclusion}
In this paper, the dual-hop mixed THz-RF cooperative relay networks have been studied. The novel closed-form expressions of the CDF and MGF for the end-to-end SNR of the considered system are derived. Further, using a well-known CDF-based approach, new closed-form expressions of generalized ASER for coherent RQAM and HQAM schemes are evaluated. The asymptotic ASER expression for RQAM scheme is also provided. Using MGF, the ASER expression of the NCFSK modulation scheme is also analyzed for the considered system. Furthermore, the impact of all considered parameters of the system is highlighted on the system's performance. All the derived expressions are validated through Monte Carlo simulation results. 

% ================================================= Appendix Section =================================================
\appendix \label{appendixSection}
This appendix elaborates on the solutions of necessary integrals required to calculate ASER for RQAM (\ref{RQAM_ASER}), ASER for HQAM (\ref{HQAM_ASER}) and asymptotic ASER for RQAM (\ref{RQAM_Asymptotic}).

% \begin{appendices}
%%%%%%%%%%% ======================== I0
\subsection{Calculation of (\ref{hop1_cdf}) from (\ref{alpha_mu_cdf}) and (\ref{channel_coeff_1_random_1_eqn})} \label{CDFISol}
With the aid of \cite{edition2002probability}, CDF of THz link $F_{h_1}\left(x\right)$ can be calculated as,
\begin{align} \label{CDFISoleq1}
    F_{h_1}\left(x\right) = \int_{y=0}^{S_0} F_{h_{f1}}\left(\frac{x}{y}\right) f_{h_{p1}}\left(y\right) dy.
\end{align}
By putting the values of $F_{h_{f1}}\left(x\right)$ and $f_{h_{p1}}\left(x\right)$ into (\ref{CDFISoleq1}) from (\ref{alpha_mu_cdf}) and (\ref{channel_coeff_1_random_1_eqn}), respectively, $F_{h_1}\left(x\right)$ can be expressed as,

\begin{align} \label{CDFISoleq2}
    F_{h_1}\left(x\right) = 1 - \frac{1}{\Gamma\left(\mu\right)}\frac{\phi}{S_0^\phi} \int_{y=0}^{S_0} y^{\phi-1} \ \Gamma\left(\mu,\frac{\mu x^\alpha}{\Omega^\alpha \ y^\alpha}\right) dy.
\end{align}
Further, with the aid of \cite{wolframIncompleteGamma}, the $\Gamma\left(\cdot,\cdot\right)$ can be written in term of Meijer G function as in (\ref{CDFISoleq3})
\begin{align} \label{CDFISoleq3}
    F_{h_1}\left(x\right) = 1 - \frac{1}{\Gamma\left(\mu\right)}\frac{\phi}{S_0^\phi} \int_{y=0}^{S_0} y^{\phi-1} \ G_{1,2}^{2,0}\left[\frac{\mu x^\alpha}{\Omega^\alpha \ y^\alpha}\left| \begin{array}{cc}
         1  \\
         \mu, 0 
    \end{array}\right.\right] dy.
\end{align}
With the aid of \cite[eq. (9.301)]{gradshteyn2014table}, the Meijer G function in (\ref{CDFISoleq3}) can be written as,

\begin{align} \label{CDFISoleq4}
    F_{h_1}\left(x\right) = 1 - &\frac{1}{\Gamma\left(\mu\right)}\frac{\phi}{S_0^\phi} \int_{y=0}^{S_0} \left(\frac{1}{2\pi i}\right) \int_s \left(\frac{\mu x^\alpha}{\Omega^\alpha S_0^\alpha}\right)^{-s} \nonumber \\
    & \times \frac{\Gamma\left(\mu+s\right)\Gamma\left(s\right)}{\Gamma\left(1+s\right)} \int_{y=0}^{S_0}y^{\phi+\alpha s -1} dy \ ds
\end{align}
After solving the inner integral, (\ref{CDFISoleq4}) can be written in term of a Meijer G function with the aid of \cite[eq. (9.301)]{gradshteyn2014table}, as expressed in (\ref{hop1_cdf}).
\begin{align} \label{CDFISoleq5}
    F_{h_1}\left(x\right) = 1 - \frac{1}{\Gamma\left(\mu\right)}\frac{\phi}{\alpha} G_{2,3}^{3,0} \left[ \frac{\mu x^\alpha}{\Omega^\alpha S_0^\alpha} \left| \begin{array}{cc}
         1, \frac{\phi}{\alpha}+1  \\
         \mu, 0, \frac{\phi}{\alpha}
    \end{array} \right.\right].
\end{align}

%%%%%%%%%%% ========================I1
\subsection{ A Solution of $\mathcal{I}_1 \left(\chi_{1}, \chi_{2} \right)$} \label{I1sol}
The integration part in (\ref{intLambdaE1}) can be written with respect to ($\chi_{2}\lambda$) as 
\begin{align} \label{intLambdaE2}
    \mathcal{I}_1 \left(\chi_{1}, \chi_{2} \right) = int_{\chi_2 \lambda = 0}^{\infty} \left(\chi_{2} \lambda\right)^{\chi_{1}} e^{-\chi_{2} \lambda} \left( \frac{1}{\chi_2}\right)^{\left(\chi_{1}+1 \right)} d\left(\chi_{2} \lambda \right).
\end{align}
From the definition of gamma function \cite[eq. (8.310.1)]{gradshteyn2014table}, (\ref{intLambdaE2}) can be written as
\begin{align} \label{intLambdaE3}
    \mathcal{I}_1 \left(\chi_{1}, \chi_{2} \right) = \left( \frac{1}{\chi_2}\right)^{\chi_1 + 1} \Gamma\left(\chi_1 + 1\right).
\end{align}
%%%%%%%%%%% ======================== I2
\subsection{A Solution of $\mathcal{I}_2 \left(\chi_1, \chi_2\right)$} \label{I2sol}
Substituting the Meijer-G representation of $\Gamma(m, \mathcal{C} \lambda)$ with the aid of \cite{wolframIncompleteGamma}, $\mathcal{I}_2\left(\cdot,\cdot\right)$ can be written as
\begin{align} \label{intLambdaEGammaGamma2}
    \mathcal{I}_2 \left(\chi_1, \chi_2\right)& = \int_{\lambda = 0}^{\infty} \frac{\lambda^{\chi_1}}{e^{\chi_2 \lambda}\, } G^{2,0}_{1,2} \left[ \mathcal{C}\lambda \left| \begin{array}{c}
    1 \\
    m, 0
    \end{array} \right. \right]  \nonumber \\
    & \times G_{2,3}^{3,0}\left[ \mu\left(\mathcal{A}\lambda\right)^{\alpha/2} \left| \begin{array}{c}
    1,\frac{\phi}{\alpha}+1 \\
    \mu,0,\frac{\phi}{\alpha}
    \end{array} \right.\right] d\lambda.
\end{align}
Further, putting the equivalent integral form of meijer G functions, the overall integral in (\ref{intLambdaEGammaGamma2}) can be expressed in term of $\mathcal{I}_1\left(\cdot,\cdot\right)$ as
\begin{align} \label{intLambdaEGammaGamma4}
    \mathcal{I}_2 \left(\chi_1, \chi_2\right) &= \left(\frac{1}{2 \pi i}\right)^2 \int_s \int_t \varphi_2\left(s\right) \varphi_3 \left(t\right)  \nonumber \\
    & \times \mathcal{I}_1\left(\chi_1-\frac{\alpha s}{2}-t, \chi_2\right) ds \ dt,
\end{align}
where, using $\Gamma\left(\chi+1\right) = \chi\Gamma\left(\chi\right)$, the functions $\varphi_2\left(s\right)$, $\varphi_3\left(t\right)$ can be expressed in simplified form as $\varphi_{2}\left(s\right) = \frac{\alpha \Gamma\left(\mu+s\right)}{s\left(\phi+\alpha s\right)}\left\{\mu\left(\mathcal{A}\right)^{\alpha/2}\right\}^{-s}$ and $\varphi_{3}\left(t\right) = \frac{\Gamma\left(m+t\right)}{t \ \mathcal{C}^t}$. Further, by putting the solution of $\mathcal{I}_1\left(\cdot,\cdot\right)$ from (\ref{intLambdaE3}), the $\mathcal{I}_2\left(\cdot,\cdot\right)$ can be expressed as a bivariate fox H function with the aid of \cite[eq. (A.1)]{Mathai2010} as
\begin{align} \label{intLambdaEGammaGamma4}
    &\mathcal{I}_2 \left(\chi_1, \chi_2\right) 
    \nonumber \\
   & 
   = \frac{1}{\chi_2^{\chi_1+1}} H_{1,0:2,3;1,2}^{0,1:3,0;2,0}\left[ \left.
    % Column 1 ========
    \begin{array}{c} 
        \mu \left(\sqrt{\frac{\mathcal{A}}{\chi_2}}\right)^{\alpha} \\
                                                               \\
        \frac{\mathcal{C}}{\chi_2}
    \end{array}
    % Column 2 ========
    \right|
    % Column 3 ========
    \begin{array}{c}
    \nu_1:\nu_2;\nu_3 \\ \\
    -:\nu_4;\nu_5
    \end{array}
    \right]
\end{align}
where, $\nu_1\left(\chi_1\right) = \left\{\left(-\chi_1;\frac{\alpha}{2},1\right)\right\}$, $\nu_2 = \left\{\left(1,1\right),\left(\frac{\phi}{\alpha}+1,1\right)\right\}$, $\nu_3 = \left\{\left(1,1\right)\right\}$, $\nu_4 = \left\{\left(\mu,1\right),\left(\frac{\phi}{\alpha},1\right),\left(0,1\right)\right\}$ and $\nu_5 = \left\{\left(m,1\right),\left(0,1\right)\right\}$.

%%%%%%%%%%% ======================== I3
\subsection{A Solution of $\mathcal{I}_3 \left(\chi_{1}, \chi_{2} \right)$} \label{I3sol}
The function ${}_1F_1(\cdot;\cdot;\cdot)$ present in (\ref{intE1F1_1}) can be expressed into its series form with the aid of \cite[eq. (9.210.1)]{gradshteyn2014table}, and therefore, the $ \mathcal{I}_3 \left(\chi_{1}, \chi_{2}, \chi_{3}\right)$ can be written as
\begin{align} \label{intE1F1_2}
    \mathcal{I}_3 \left(\chi_{1}, \chi_{2}, \chi_{3}\right) = \sum_{\eta=0}^{\infty} \frac{\Gamma\left(\frac{3}{2}\right)\Gamma\left(\eta+1\right)}{\Gamma\left(\frac{3}{2}+\eta\right)}\frac{\chi_3^\eta}{\eta!}\int_{\lambda = 0}^{\infty} \lambda^{\eta+\chi_1} e^{-\chi_{2} \lambda} d\lambda.
\end{align}
The integration part in (\ref{intE1F1_2}) is similar to $\mathcal{I}_{1}(\cdot,\cdot)$ and thus can be solved by using (\ref{intLambdaE3}) as
\begin{align} \label{intE1F1_3}
    \mathcal{I}_3 & \left(\chi_{1}, \chi_{2}, \chi_{3}\right) \nonumber \\
    & = \sum_{\eta=0}^{\infty} \frac{\Gamma\left(\frac{3}{2}\right)\Gamma\left(\eta+1\right)}{\Gamma\left(\frac{3}{2}+\eta\right)}\frac{\chi_3^\eta}{\eta!} \mathcal{I}_{1}\left(\eta+\chi_1,\chi_2\right) \nonumber \\
    & = \frac{\Gamma\left(\chi_1+1\right)}{\chi_2^{\chi_1+1}} \sum_{\eta=0}^{\infty} \frac{\Gamma\left(\eta+1\right)\Gamma\left(\frac{3}{2}\right)\Gamma\left(\eta+1\right)}{\Gamma\left(\frac{3}{2}+\eta\right)} \frac{\left(\chi_3/\chi_2\right)^\eta}{\eta!}.
\end{align}
Further, with the aid of \cite[eq. (7.2.3)]{prudnikov}, the $ \mathcal{I}_3 \left(\chi_{1}, \chi_{2}, \chi_{3}\right)$ can be written in term of hypergeometric function as
\begin{align} \label{intE1F1_4}\textbf{}
    \mathcal{I}_3 \left(\chi_{1}, \chi_{2}, \chi_{3}\right) = \frac{\Gamma\left(\chi_1+1\right)}{\chi_2^{\chi_1+1}} \, {}_{2}F_{1}\left(1,1;\frac{3}{2};\frac{\chi_3}{\chi_2}\right).
\end{align}

%%%%%%%%%%% ======================== I4
\subsection{A Solution of $\mathcal{I}_4 \left(\chi_1, \chi_2 \right) $} \label{I4sol}
The function ${}_1F_1(\cdot;\cdot;\cdot)$ and $\Gamma(\cdot,\cdot)$ present in (\ref{intLambdaEGammaGamma1F1_1}) can be expressed into their equivalent meijer G representation with the aid of \cite{wolframKummerConfluent} and \cite{wolframIncompleteGamma} respectively as
\begin{align} \label{intLambdaEGammaGamma1F1_2}
    \mathcal{I}_4 &\left(\chi_1, \chi_2 \right) = \frac{1}{2}\int_{\lambda=0}^{\infty} e^{-\left(\chi_1-\chi_2\right)\lambda} G_{1,2}^{1,1}\left[\chi_2\lambda\left|\begin{array}{c} 
        \frac{1}{2}     \\
        0, -\frac{1}{2} \\
    \end{array}\right.\right]  \nonumber \\
    & \times G_{2,3}^{3,0}\left[\mu\left(\mathcal{A}\lambda\right)^{\alpha/2}\left|\begin{array}{c} 
        1, \frac{\phi}{\alpha}+1     \\
        \mu, 0, \frac{\phi}{\alpha} \\
    \end{array}\right.\right]G_{1,2}^{2,0}\left[\mathcal{C}\lambda\left|\begin{array}{c} 
        1  \\
        m, 0 \\
    \end{array}\right.\right] d\lambda.    
\end{align}
Further, with the aid of [Definition of Meijer G], meijer G functions in (\ref{intLambdaEGammaGamma1F1_2}) can be expressed with their equivalent integral form. Thus, after performing some simplifications using $\Gamma\left(\chi+1\right) = \chi\Gamma\left(\chi\right)$, the overall integral can be written in term of $\mathcal{I}_{1}\left(\cdot,\cdot\right)$ as
\begin{align} \label{intLambdaEGammaGamma1F1_3}
    \mathcal{I}_4 \left(\chi_1, \chi_2 \right) = &\left(\frac{1}{2\pi i}\right)^3 \int_r \int_s \int_t \varphi_1\left(r\right) \varphi_2\left(s\right) \varphi_3\left(t\right)\nonumber \\
    &  \times \mathcal{I}_1\left(-r-\frac{\alpha s}{2}-t, \chi_1-\chi_2\right) dr \ ds \ dt,
\end{align}
where, $\varphi_{1}\left(r\right) = \frac{\Gamma\left(r\right)}{\left(1-2r\right) \ \chi_2^r}$. After solving the integral $\mathcal{I}_1\left(\cdot, \cdot\right)$, equation (\ref{intLambdaEGammaGamma1F1_3}) can be written as
%\textcolor{red}{By using the Mellin transform of the Product of two H-functions \cite[eq. (2.25.1.1)]{prudnikov}, the integration in (\ref{intLambdaEGammaGamma3}) can be solved  as
\begin{align} \label{intLambdaEGammaGamma4}
    \mathcal{I}_4 \left(\chi_1, \chi_2 \right) = &\left(\frac{1}{2\pi i}\right)^3 \int_r \int_s \int_t \varphi_1\left(r\right) \varphi_2\left(s\right) \varphi_3\left(t\right)  \nonumber \\
    & \times\frac{\Gamma\left(1-r-\frac{\alpha s}{2}-t\right)}{\left(\chi_1 - \chi_2\right)^{1-r-\left(\alpha s/2\right)-t}} dr \ ds \ dt.
\end{align}
Further, with the aid of \cite[eq. (A.1)]{Mathai2010}, equation (\ref{intLambdaEGammaGamma4}) can be written in term of trivariate fox H as
%Equation (\ref{intLambdaEGammaGamma4}) can be written in terms of $\Psi_1(\cdot,\cdot,\cdot)$ as
\begin{align} \label{intLambdaEGammaGamma5}
    &\mathcal{I}_4 \left(\chi_1, \chi_2\right) = \frac{1}{2\left(\chi_1-\chi_2\right)} 
   \nonumber \\
    &\times H_{1,0:1,2;2,3;1,2}^{0,1:1,0;3,0;2,0}\left[ 
    \left.\begin{array}{c} 
        \left(\frac{\chi_2}{\chi_1-\chi_2}\right)  \\ \\
        \mu\left(\frac{\mathcal{A}}{\chi_1-\chi_2}\right)^{\alpha/2} \\ \\
        \left(\frac{\mathcal{C}}{\chi_1-\chi_2}\right)
    \end{array}
    \right|
    \begin{array}{c} 
        \nu_6:\nu_7;\nu_2;\nu_3 \\
        -:\nu_8;\nu_4;\nu_5
    \end{array}
    \right],
\end{align}
where, $\nu_6 = \left\{\left(0;1,\frac{\alpha}{2},1\right)\right\}$, $\nu_7 = \left\{\left(\frac{1}{2},1\right)\right\}$ and $\nu_8 = \left\{\left(0,1\right), \left(-\frac{1}{2}\right)\right\}$.

\bibliographystyle{IEEEtran}
% \bibliography{IEEEabrv, bibliography}
\bibliography{bibliography}

\end{document}